\documentclass[sigconf, nonacm]{acmart}
\acmConference[EASE 2026]{The 30th International Conference on Evaluation and Assessment in Software Engineering}{9–12 June, 2026}{Glasgow, Scotland, United Kingdom}
\usepackage{tabularx}
\usepackage{makecell}
\usepackage{float} 
\usepackage{listings}
\usepackage{hyperref}
\usepackage{framed}
\usepackage{setspace}
\usepackage{tcolorbox}
\usepackage{float}
\usepackage{color}
\definecolor{mygray}{gray}{.9}

\tcbuselibrary{skins, breakable, theorems, hooks}

\usepackage[ruled,vlined,linesnumbered]{algorithm2e}
\usepackage{hyperref}
\hypersetup{
    colorlinks=true, 
    urlcolor=magenta,   
}

\newtcolorbox{mybox}[2][]{
top=0.15in,left=4pt,right=4pt,bottom=4pt,
fonttitle=\bfseries,
colbacktitle=gray,
colback=gray!5,
colframe=gray!40!black,
enhanced,
attach boxed title to top left={xshift=1.5em,yshift=-\tcboxedtitleheight/2},
boxed title style={size=small},
drop shadow={black!50!white},
title=#2,#1}

\AtBeginDocument{%
  }

\setcopyright{acmlicensed}
\copyrightyear{2026}
\acmYear{2026}
\acmDOI{XXXXXXX.XXXXXXX}

\acmISBN{978-1-4503-XXXX-X/2018/06}

\begin{document}

\title{Boosting Automatic Java-to-Cangjie Translation with Multi-Stage LLM Training and Error Repair}


\author{Xinyue Liang}
\email{xy_liang@nuaa.edu.cn}
\affiliation{
  \institution{Nanjing University of Aeronautics and Astronautics}
  \country{China}
}

\author{Jingxuan Zhang}
\email{jxzhang@nuaa.edu.cn}
\authornote{Jingxuan Zhang is the corresponding author.}
\affiliation{
  \institution{Nanjing University of Aeronautics and Astronautics}
  \country{China}
}

\author{Lin Li}
\email{this_is_lilin@nuaa.edu.cn}
\affiliation{
  \institution{Nanjing University of Aeronautics and Astronautics}
  \country{China}
}
\author{Jun Zhang}
\email{zhang_jun@nuaa.edu.cn}
\affiliation{
  \institution{Nanjing University of Aeronautics and Astronautics}
  \country{China}
}
\author{Junhao Chen}
\email{chenjjhh@nuaa.edu.cn}
\affiliation{
  \institution{Nanjing University of Aeronautics and Astronautics}
  \country{China}
}








\begin{abstract}
With the rapid evolution of emerging programming language ecosystems, the demand for code translation to low-resource languages continues to grow. As Cangjie emerges as a new programming language, its ecosystem and development toolchains are rapidly expanding. Automated translation from popular programming languages to Cangjie is therefore valuable for practical development. However, constrained by both insufficient Cangjie knowledge and scarce parallel code corpora, general Large Language Models (LLMs) are prone to syntactic errors and semantic as well as structural misalignment in code translation. Existing approaches typically rely on fine-tuning with large-scale parallel data, but they cannot reliably improve compilability or semantic consistency for low-resource Cangjie languages. To tackle these challenges, we propose a multi-stage training framework of LLMs that employs the iterative error repair technique to translate Java code into Cangjie code. This training framework performs training on LLMs, gradually integrating knowledge and achieving semantic alignment as well as structure awareness. During the code translation, we also combine the compiler feedback and error repair case retrieval to repair the incorrect Cangjie code. We construct syntactic knowledge and monolingual instruction datasets to train the LLM. In addition, we also build a Cangjie error repair repository to support error repair in our approach. Experimental results show that, with limited parallel data, our approach improves functional equivalence by 6.06\% compared to the state-of-the-art approaches. Meanwhile, ablation studies confirm that each training stage positively contributes to the final performance.
\end{abstract}





\maketitle

\section{Introduction}

As software systems continue to grow in scale, the evolution of programming languages and runtime platforms has become common in software engineering practice. Enterprises frequently migrate existing code across languages to address ecosystem transitions or cross-platform compatibility \cite{migrate00_2021,migrate01_2022,migrate02_2022}. To reduce manual costs and improve migration efficiency, automated code translation is emerging. It can accelerate the continuous delivery and long-term maintenance of new platforms and ecosystems, improving system sustainability \cite{migrate03_2025}. Automated code translation generates code that not only needs to be syntactically correct, but also needs to be able to compile, run, and maintain functional consistency. However, existing approaches struggle to achieve this, especially for low-resource languages \cite{howfar_2024}.

This paper studies the automated code translation from Java to Cangjie. Cangjie is an emerging programming language in the HarmonyOS ecosystem. Cangjie is an emerging multi-paradigm programming language designed for native HarmonyOS applications. In contrast, Java has a mature ecosystem and a large scale codebase, and it is widely used in enterprise software. High-quality Java-to-Cangjie translation has a strong demand in practice. 

In the literature, researchers have proposed several code translation approaches. Early code translation approaches primarily relied on heuristic rules \cite{c2rust,cxgo,java2csharp}. These translators performed pattern matching and rewriting on Abstract Syntax Trees (ASTs). They required domain experts to propose and maintain heuristic rules, which led to high development and maintenance costs. In addition, the generated code also often failed to follow target-language idioms, so manual correction was still needed \cite{ruleBad_2013}. Later, statistical machine translation and neural-based translation were applied to code translation \cite{statis_DaC_2015,neural_Tree_2018,roziere_2022}. They learned cross-language mappings from parallel code corpora and improved performance on several programming language pairs. However, these approaches depend on high-quality parallel code corpora. In recent years, Large Language Models (LLMs) pre-trained on large code and text corpora have shown strong generalization ability and have been applied to code translation. Compared with models trained mainly on parallel code corpora, LLMs can generate readable target code in zero-shot or few-shot settings \cite{codetransocean_2023}. They can also interpret natural language to capture developers' intent better \cite{cotran_2024}. However, existing studies show that LLMs still struggle to guarantee the correctness of the translation. Pan et al. \cite{lost_2024} evaluated multiple code translators on real-world projects. They reported that the average translation success rate ranged from 2.1\% to 47.3\% for LLMs. They also summarized several defect types, which suggests that fully reliable automation of code translation is not yet achieved. To improve the translation success rate, later work incorporated the compiler feedback and test execution information into the code translation. For example, UniTrans \cite{unitrans_2024} leverages automatically generated unit tests to augment translation prompts and uses execution feedback to correct faulty translations. TransAGENT \cite{transagent_2024} frames LLM-based code translation as a collaborative multi-agent workflow. They improve the performance of code translation by using different agents to handle syntactic errors and semantic discrepancies.

Despite these advances, code translation to low-resource programming languages still remains under-studied. First, LLMs have insufficient pre-training coverage of low-resource programming languages \cite{lowresLLM_2024}. Insufficient knowledge of syntax and standard libraries often leads to compilation errors and semantic deviations for LLMs \cite{lost_2024}. Second, the available parallel data for low-resource programming languages is limited in scale. Many studies rely on the back-translation approach or the synthetic data to augment training sets \cite{Roziere_2020}. However, the synthetic data may contain noise and bias \cite{roziereST_2021,NoDataAug_2024}. Therefore, code translation targeted towards low-resource programming languages requires addressing both the knowledge gap and the data scarcity problems.


This study proposes a Java-to-Cangjie translation approach with multi-stage LLM training and error repair. The goal is to improve syntactic correctness, semantic consistency, and functional consistency for Java-to-Cangjie translation under the data scarcity constraint. We train an LLM in multiple stages. The LLM first learns the structured syntactic knowledge of Cangjie, then progressively aligns semantics and code structure through fine-tuning, and finally conducts iterative error repair with LLMs and Retrieval Augmented Generation (RAG). Building on this overall pipeline, the paper develops a multi-stage training and repair framework comprising four stages. (i) Continued pre-training (CPT) for the LLM is conducted on structured syntactic knowledge to inject Cangjie syntax and key APIs. (ii) Semantic monolingual instruction fine-tuning for the LLM learns a robust mapping from natural-language to Cangjie code under limited data scale. (iii) An AST structure-aware mechanism for the LLM maps Java structures into trainable structure tokens, and parallel fine-tuning strengthens the structural constraints during the translation. (iv) An iterative error repair process combines LLM-based self-analysis repair and RAG enhanced repair to support iterative translation and correction.

We conducted a series of experiments to evaluate the effectiveness of the approach based on a public Java-to-Cangjie benchmark dataset \cite{cjTrans_2025}. We design several evaluation metrics, including functional equivalence, compilation success rate, text similarity, and compilable conditional functional equivalence. The approach achieves 63.64\% functional equivalence and improves by 6.06\% over the state-of-the-art baseline. This indicates that our approach improves functional correctness under the low-resource scenario. We also conduct ablation studies to evaluate the contribution of each stage. Removing any stage reduces functional equivalence by 3.63\% to 7.27\%, which shows that each stage contributes to the final result. Furthermore, we compare the effectiveness of different error repair strategies. Under the same training setting, LLM-based self-analysis repair and RAG enhanced repair achieve 56.97\% and 61.21\% functional equivalence. We combine LLM-based self-analysis repair with RAG-enhanced repair, the functional equivalence further increases to 63.64\%, which confirms the advantage of integrating the two repair strategies. We further evaluate the performance of different LLMs. Our approach yields consistent gains across different LLMs, indicating good transferability.

In summary, the main contributions of this paper are as follow:

\begin{enumerate}
    \item We propose a multi-stage training framework for LLM to perform low-resource code translation. The framework combines knowledge integration, semantic alignment, and structure awareness to improve the translation effectiveness from Java-to-Cangjie.
    \item We design an iterative error repair process that integrates LLM-based self-analysis repair and RAG enhanced repair. By constructing an error repair repository for the Cangjie language, we improve the compilability and functional consistency of the generated code.
    \item We conduct extensive experiments to validate the performance of our approach. In addition, we open the dataset and replication package to the public to facilitate subsequent reproduction and extension\footnote{\url{https://github.com/lionqoo/translator}}. 
\end{enumerate}

\section{Motivation}

\subsection{Challenges in Low-Resource Code Translation}

Translating widely used Java code into the emerging Cangjie language is important for cross-platform migration. However, Cangjie is still under development, and high-quality Java-to-Cangjie parallel corpora are scarce. This low-resource scenario limits the effectiveness of traditional code translation approaches \cite{lowres_2023}. This challenge is exacerbated by the knowledge imbalance in LLMs. Most LLMs are pre-trained mainly on high-resource programming languages such as Java and Python. As a result, they often lack sufficient knowledge of the core syntax, the usage of standard library API, and runtime behavior in low-resource programming languages, e.g., Cangjie \cite{lowresLLM_2024}.

\begin{figure}[t]
    \centering
    \includegraphics[width=1\linewidth]{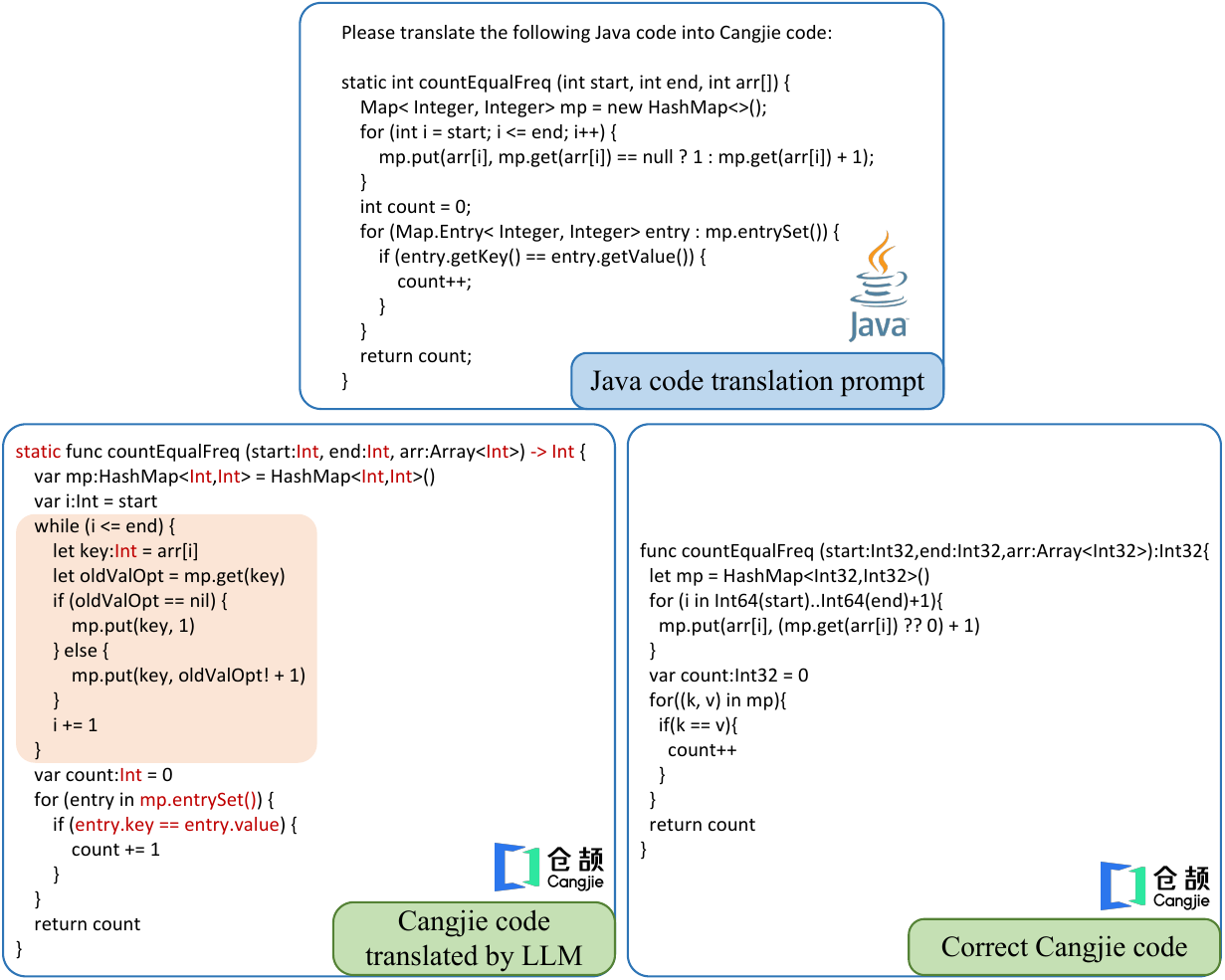}
    \caption{Comparisons Between LLM-Generated and Reference Cangjie Translations.}
    \label{fig:Comparison}
\end{figure}

Figure \ref{fig:Comparison} shows the code translation results generated by GPT-5.2 when directly translating the Java code into Cangjie without task-specific adaptation. We compare the output with correct one. The red annotations indicate the syntax-level errors in the code translation result, and the orange boxes indicate fragments that deviate from the source program structure. We observe that the output often contains type mismatches (e.g., generalizing what should be Int32 numeric types to Int). It also incorrectly transfer Java-style container iteration to Cangjie (e.g., migrating Java entrySet patterns to Cangjie). These issues can cause compilation failures or semantic deviations. Therefore, LLMs often fail to produce code that conforms to low-resource language specifications.

In this scenario, supervised learning approaches that rely on large-scale parallel data have significant limitations. In particular, the existing Cangjie parallel corpora is limited in both size and coverage. Therefore, these corpora cannot support stable training when relying solely on CPT or using only a small amount of parallel data for supervised fine-tuning \cite{cptSurvey_2024}. In addition, many code translation approaches using LLMs adopt iterative repair driven by compilation or execution feedback \cite{cotran_2023}, but in low-resource target-language scenario, relying only on LLMs to generate fixes from compiler errors often leads to the following problems: (i) For unseen syntactic structures, LLMs often fail to produce effective corrections. (ii) When fixing errors based solely on compiler error messages, the generated fixes are difficult to generalize to semantically different code snippets \cite{fixBug1_2017,fixBug2_2022}. 



In summary, the main difficulty is not only the limited target-language knowledge of LLMs. A more fundamental challenge is the lack of large-scale parallel corpora. It is necessary to construct an efficient and scalable training and repair pipeline for LLMs.

\subsection{Principles of the Proposed Approach}

Based on the above discussion, our research motivation includes two aspects. In the training phase, we employ a multi-stage training pipeline to establish target-language proficiency and cross-lingual alignment under the low-resource scenario. In the inference phase, we use an iterative error repair process that combines LLM-based self-analysis repair and RAG enhanced repair to improve translation ability.

With limited parallel corpora, fine-tuning on LLMs using small-scale parallel code corpora often fails to deliver stable improvements \cite{dataAug_2023}. To address this, we draw inspiration from human learning methods and adopt a staged training pipeline. This process begins with knowledge integration, proceeds through semantic modeling, and then performs structural mapping, thereby making full use of the effective information within limited resources. 

\begin{figure*}[h]
    \centering
    \includegraphics[width=0.9\linewidth]{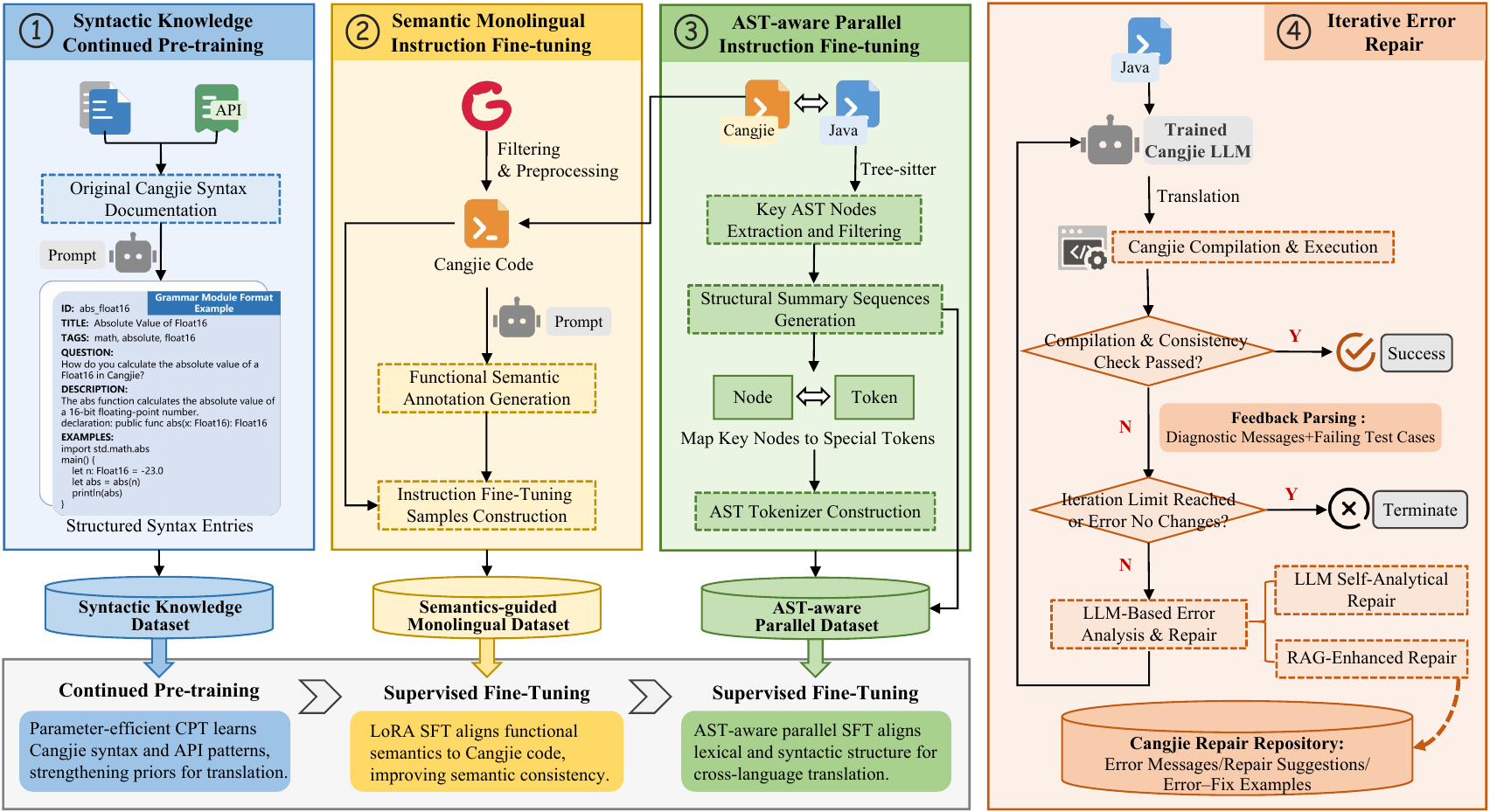}
    \caption{Overall Framework of the Proposed Java-to-Cangjie Translation Pipeline.}
    \label{fig:main}
\end{figure*}

Although the training stage can enhance target-language knowledge and structural alignment capabilities, low-resource translation may still produce compilation errors and functional deviations \cite{lowresLLM_2024,motivate_2025}. To address this, we design a repair process combining LLM-based self-analysis repair and RAG enhanced repair during the inference stage. The process links compilation errors to an error repair repository. In each round, it retrieves validated error repair examples as external guidance. This improves repair accuracy and generalization.

In summary, for low-resource code translation, we propose a multi-stage training framework for LLM with iterative error repair during inference. 

\section{Approach}

Figure \ref{fig:main} shows our multi-stage training and iterative repair pipeline. It consists of syntactic knowledge pre-training, semantic monolingual instruction fine-tuning, AST-aware parallel fine-tuning, and inference-time iterative error repair. Details are introduced as follows.


\subsection{Syntactic Knowledge Pre-training}

To address the issue of insufficient language knowledge coverage in Cangjie, this stage constructs Cangjie prior knowledge for the LLM. We reconstruct the Cangjie type constraints, syntactic templates, and standard documentation database into a structured corpora. The corpora are then injected into the LLM through CPT. This process provides the syntactic foundation for the following stages.

\begin{figure}
    \centering
    \includegraphics[width=1\linewidth]{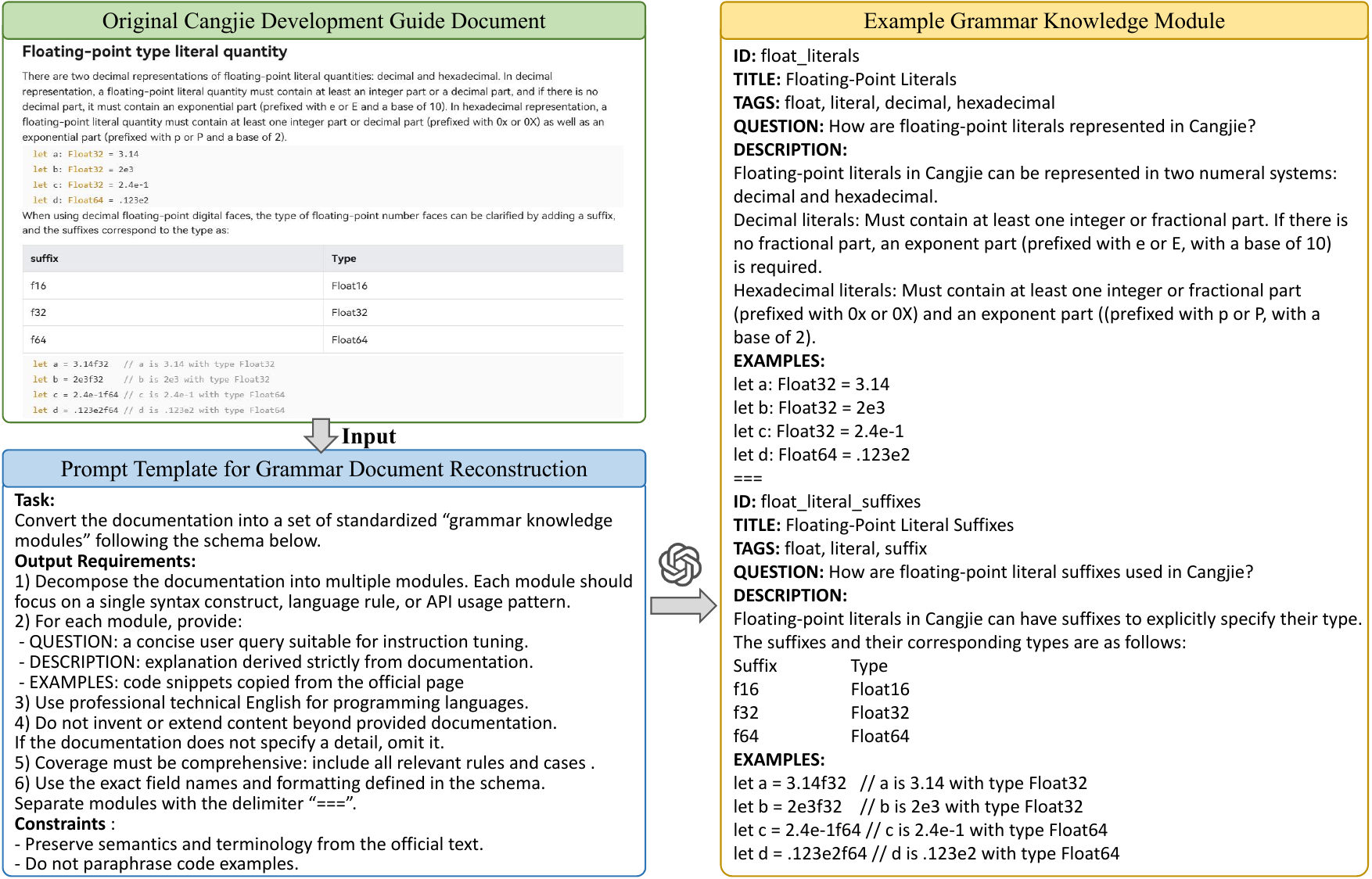}
    \caption{Cangjie Documentation Reconstruction Prompt.}
    \label{fig:document}
\end{figure}

\textbf{Syntax Documentation Reconstruction:} The original Cangjie syntax documentation is an official developer guide \footnote{\url{https://cangjie-lang.cn/docs}}. It is organized by different chapters and includes natural language descriptions, lists, tables, code examples, and comments. However, the content is mostly unstructured and inconsistent across different chapters. It is difficult to directly convert the documentation into training samples for LLMs. To address this issue, we employ an LLM-based reconstruction pipeline. It transforms the Cangjie syntax documentation into semantically consistent and uniformly formatted entries. We first segment the documentation by chapters and design reconstruction prompt templates. Each chapter is then processed by an LLM together with the prompt template to generate structured entries. Figure \ref{fig:document} shows the prompt template and examples of the resulting modules. Each entry contains a unique ID, title, tags, typical questions, normalized descriptions, and executable code examples. The entries cover both concept definitions and usage patterns of Cangjie syntax. 

\textbf{Data Configuration:} These entries are transformed into text formats suitable for LLM pre-training. We concatenate the entry fields in a fixed order and insert uniform paragraph boundary markers. This design keeps a stable internal structure for each entry. We executed knowledge extraction and processing on 122 chapters in the Cangjie syntax documentation, with multiple rounds of manual verification. This process produced a syntactic knowledge dataset with 6779 entries and over 3400 core semantic concepts.

\textbf{Continued Pre-training:} We apply CPT to the LLM using the syntactic knowledge dataset. The LLM learns Cangjie syntax rules, key structures, and standard API usage patterns through token-by-token prediction. This pre-training provides stronger priors for instruction fine-tuning and cross-language translation. To inject new knowledge and maintain training stability, we adopt the parameter-efficient LoRA \cite{lora_2022} technique. Most parameters are frozen, and only low-rank adaptation parameters are updated. This strategy can reduce memory and computation costs and limit interference with the distribution of the LLM. It also supports a smoother transition from a general code distribution to a Cangjie-specific syntax distribution. 

\subsection{Semantic Monolingual Instruction Fine-tuning}

This stage utilizes Cangjie monolingual code to construct a dataset with explicit functional semantic annotations, followed by instruction fine-tuning. The goal of this stage is to learn mappings between functional descriptions and code implementations for Cangjie.

\textbf{Data Collection:} We collect the source code from the Cangjie open-source community in GitCode\footnote{ \url{https://gitcode.com}}. Given the limited community size, to ensure code quality and implementation reliability, we prioritized code repositories with at least 50 stars as candidate data sources. These repositories were decomposed into candidate code snippets. During filtering, we excluded code snippets with fewer than 5 lines. To minimize the impact of external dependencies, we retained the code snippets with fewer third-party dependencies and clearly defined functionalities. Specifically, the selected snippets primarily utilize the Cangjie standard library and avoid excessive calls to external frameworks. Furthermore, we retained code snippets that are complete in isolation, excluding cross-module code and those with incomplete control flow. To enrich this dataset, Cangjie code snippets from the parallel translation data used in the AST-aware training stage are also added to the monolingual collection. This inclusion increases data coverage and creates partially shared target-language samples between training stages.

\textbf{Generating Semantic Annotations:} Although generating correct code in low-resource programming languages remains challenging for LLMs, these LLMs possess robust capabilities in code understanding and semantic abstraction. These abilities can also transfer to low-resources programming languages like Cangjie \cite{LLMsemantic_2025,LLMsemantic2_2024}. Therefore, we use DeepSeek-V3.2 to generate function-level semantic descriptions for each Cangjie code snippet. During generation, the function signature, function body, and necessary local comments for each code snippet are provided as input. DeepSeek-V3.2 is required to output a concise functional description without adding external context. The prompt template for DeepSeek-V3.2 is as follows:

\begin{lstlisting}
### You are an assistant for code semantic interpretation.
### Summarize the functional semantics of the following Cangjie code in one concise, imperative-style natural language sentence. Output only the description.
[Cangjie_Code]
\end{lstlisting}

To mitigate the noise generated by DeepSeek-V3.2 and reduce semantic drift, the generated functional descriptions were manually evaluated by two annotators with Cangjie programming experience. We randomly selected 100 samples and asked the annotators to verify the alignment between the code and its description. The evaluation resulted in 96\% accuracy, demonstrating that the LLM can generate reliable functional descriptions for Cangjie code snippets.
This random sampling serves as a statistical quality check rather than a full labeling process, and the high accuracy supports using the generated annotations to scale the dataset under limited expert resources.

\textbf{Dataset Construction:} We structure the generated functional descriptions and the Cangjie code snippets into unified samples for instruction fine-tuning. For the purpose of the code generation task, we assign the roles of input and output to establish a text-to-code mapping: the functional description serves as the input, while the Cangjie code serves as the target output. The instruction specifies the task of generating code from the description. Through this process, we constructed 3,241 monolingual samples, effectively transforming the raw code base into a semantic-guided dataset.

\textbf{Semantic Instruction Fine-tuning:} We utilized the LLM from the previous stage and conducted Supervised Fine-Tuning (SFT) on the monolingual dataset. A parameter-efficient LoRA configuration is employed to preserve existing Cangjie knowledge and maintain training stability. In this stage, our approach learns to generate syntactically correct and functionally consistent Cangjie code from natural language functional descriptions. This training establishes explicit links between functional semantics and Cangjie implementations. Compared with training based solely on general code modeling, our SFT-based approach focuses more on semantic alignment and standardized Cangjie generation patterns. This focus enhances compilation correctness and functional consistency in subsequent cross-language translation. After this stage, our approach retains Cangjie fundamentals and strengthens the mapping from functional semantics to code implementations.

\subsection{AST-aware Parallel Instruction Fine-tuning}

This stage mitigates structural deviation issues in code translation by introducing the source language AST as constraints. The primary objective is to enhance both the structural consistency and the correctness of the translated code. The process comprises three main steps, including extracting structural summary sequences from Java source code, structural token design, and parallel instruction fine-tuning. These steps enable our approach to incorporate both the source-language code and its AST structure when generating Cangjie code. 

\textbf{Generating Structural Summary Sequences:} We construct concise and structural summary sequences in this step. We parse Java code with Tree-sitter \cite{treesitter} and apply a depth-first traversal to linearize the AST, retaining only control-flow and semantic node types. During the traversal, we preserve node types related to function semantics and control flow, such as method declarations, parameter structures, conditional branches, and loop structures. In addition, we discard terminal nodes, which are mostly token-level elements, and retain only internal structural nodes. This step produces a compact structural summary sequence for each AST. It preserves the main control-structure information and reduces redundancy. This step facilitates the integration of structural data alongside source code within the limits of context length.

\textbf{Structural Token Design:} To incorporate the structural information, we discretize the structural summaries into structural tokens and extend the tokenizer vocabulary accordingly. We categorize Java AST nodes based on their structural semantics (e.g., control flow) and map each category to a special token. The embeddings of these tokens are trainable parameters updated during training, so structural signals are encoded directly in the embedding space. This enables the transformer to attend to both lexical and structural tokens simultaneously, thereby improving structural consistency across languages.
\begin{figure}
    \centering
    \includegraphics[width=1\linewidth]{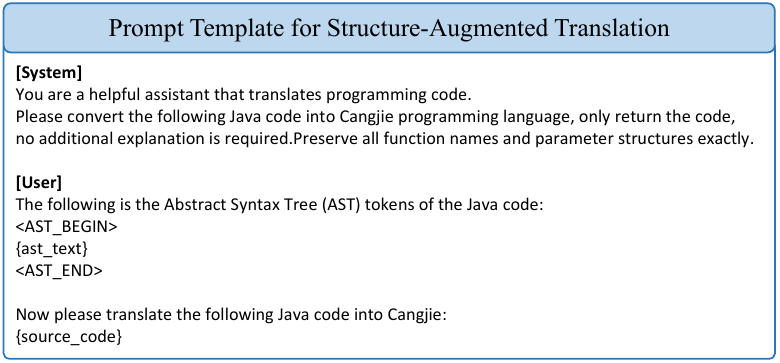}
    \caption{AST-Conditioned Prompt Template for Structure-Augmented Java-to-Cangjie Translation.}
    \label{fig:ast-conditioned}
\end{figure}

\textbf{Parallel Instruction Fine-tuning:} Building on the above steps, we integrate structural signals with translation tasks through AST-aware parallel SFT. We use a unified prompt template, as shown in Figure \ref{fig:ast-conditioned}, which integrates structural summary sequences, the source code, and task instructions within a single input. To resolve ambiguity, we insert boundary markers that explicitly demarcate structural snippets from code. Subsequently, we employ LoRA for parameter-efficient SFT. By leveraging this method, we align the embeddings of these new tokens with translation objectives, thereby minimizing training costs while ensuring effective adaptation. While the overall methodology maintains an auto-regressive generation objective, it explicitly imposes structural constraints on the input side. This enables the LLM to align semantic content with program structure when learning cross-language mappings.


By implementing AST-aware parallel fine-tuning, our approach can consider lexical sequences and syntactic structure at the same time, thereby reducing structural errors such as control flow structure deviations. It also improves the compilation rate and structural consistency of the translated results, which provides a critical foundation for generating correct Cangjie code.

\subsection{Iterative Error Repair}

To enhance the execution accuracy of the generated Cangjie code, we propose an iterative error repair framework. This automated pipeline integrates compilation and testing feedback to continuously refine and rectify the translated code. Algorithm 1 outlines the detailed procedure. Initially, our approach compiles and tests the translated Cangjie code (lines 2-3). When detecting compiler errors or testing failures, our approach triggers the error repair mechanism (line 8). The error repair mechanism dynamically chooses between RAG-enhanced repair and LLM-based self-analysis to repair the compiler errors and testing failures (line 9). If the similarity score exceeds a predefined threshold, our approach leverages RAG to retrieve similar error repair cases in the error repair repository, utilizing them to guide the error repair process. (lines 10-11). Conversely, if the similarity score falls below the threshold or if the errors are limited to functional inconsistencies, our approach switches to LLM-based self-analysis repair (lines 12-17). After the error repair, the new version of the Cangjie code is generated, compiled, and tested again. The iterative error repair procedure terminates either when it reaches the maximum number of iterations or when the error pattern no longer changes. 
This design allows our approach to dynamically select between RAG-enhanced repair and LLM-based self-analysis, optimizing repair efficiency while maintaining broad applicability.

\begin{algorithm}[t]\scriptsize
\caption{Iterative Error Repair Algorithm}
\SetAlgoLined
\DontPrintSemicolon

\SetKwInput{KwIn}{Input}
\SetKwInput{KwOut}{Output}

\SetKwFor{While}{while}{do}{}
\SetKwIF{If}{ElseIf}{Else}{if}{then}{else if}{else}{}

\KwIn{Source Java code $J$, Test case set $T$, Error case library $D$, Similarity threshold $\tau$, Maximum iteration number $N_{max}$}
\KwOut{Cangjie code after the $k$-th translation $C^{(k)}$}

\While{$k < N_{max}$}{
    $(\mathit{compileStatus}, \mathit{errorMsg}) \leftarrow \mathrm{Compile}(C^{(k)})$\;
    $\mathit{testResult} \leftarrow \mathrm{Test}(C^{(k)}, T)$\;

    \If{$\mathit{compileStatus}=\mathit{success}$ \textbf{and} $\mathit{testResult}=\mathit{pass}$}{
        \Return $C^{(k)}$ as the final result\;
    }
    \ElseIf{the error no longer changes}{
        \Return $C^{(k)}$\;
    }
    \ElseIf{$\mathit{compileStatus}=\mathit{fail}$}{
        $(\mathit{similarCases}, \mathit{similarScore}) \leftarrow \mathrm{Retrieve}(\mathit{errorMsg}, D)$\;
        \If{$\mathit{similarScore}\ge \tau$}{
            $C^{(k+1)} \leftarrow \mathrm{LLM}\!\big(\mathrm{RAG\text{-}Prompt}(\mathit{errorMsg}, \mathit{similarCases}, C^{(k)})\big)$\;
        }
        \Else{
            $G_{\mathit{rep}} \leftarrow \mathrm{LLM}\!\big(\mathrm{Repair\text{-}Prompt}(\mathit{errorMsg}, J, C^{(k)})\big)$\;
            $C^{(k+1)} \leftarrow \mathrm{LLM}\!\big(\mathrm{Repair\text{-}Prompt}(\mathit{errorMsg}, G_{\mathit{rep}}, J, C^{(k)})\big)$\;
        }
    }
    \ElseIf{$\mathit{compileStatus}=\mathit{success}$ \textbf{and} $\mathit{testResult}=\mathit{fail}$}{
        $G_{\mathit{rep}} \leftarrow \mathrm{LLM}\!\big(\mathrm{Repair\text{-}Prompt}(\mathit{errorCases}, J, C^{(k)}, \mathit{Output})\big)$\;
        $C^{(k+1)} \leftarrow \mathrm{LLM}\!\big(\mathrm{Repair\text{-}Prompt}(\mathit{errorCases}, G_{\mathit{rep}}, J, C^{(k)})\big)$\;
    }

    $k \leftarrow k + 1$\;
}
\end{algorithm}

\begin{figure}
    \centering
    \includegraphics[width=1\linewidth]{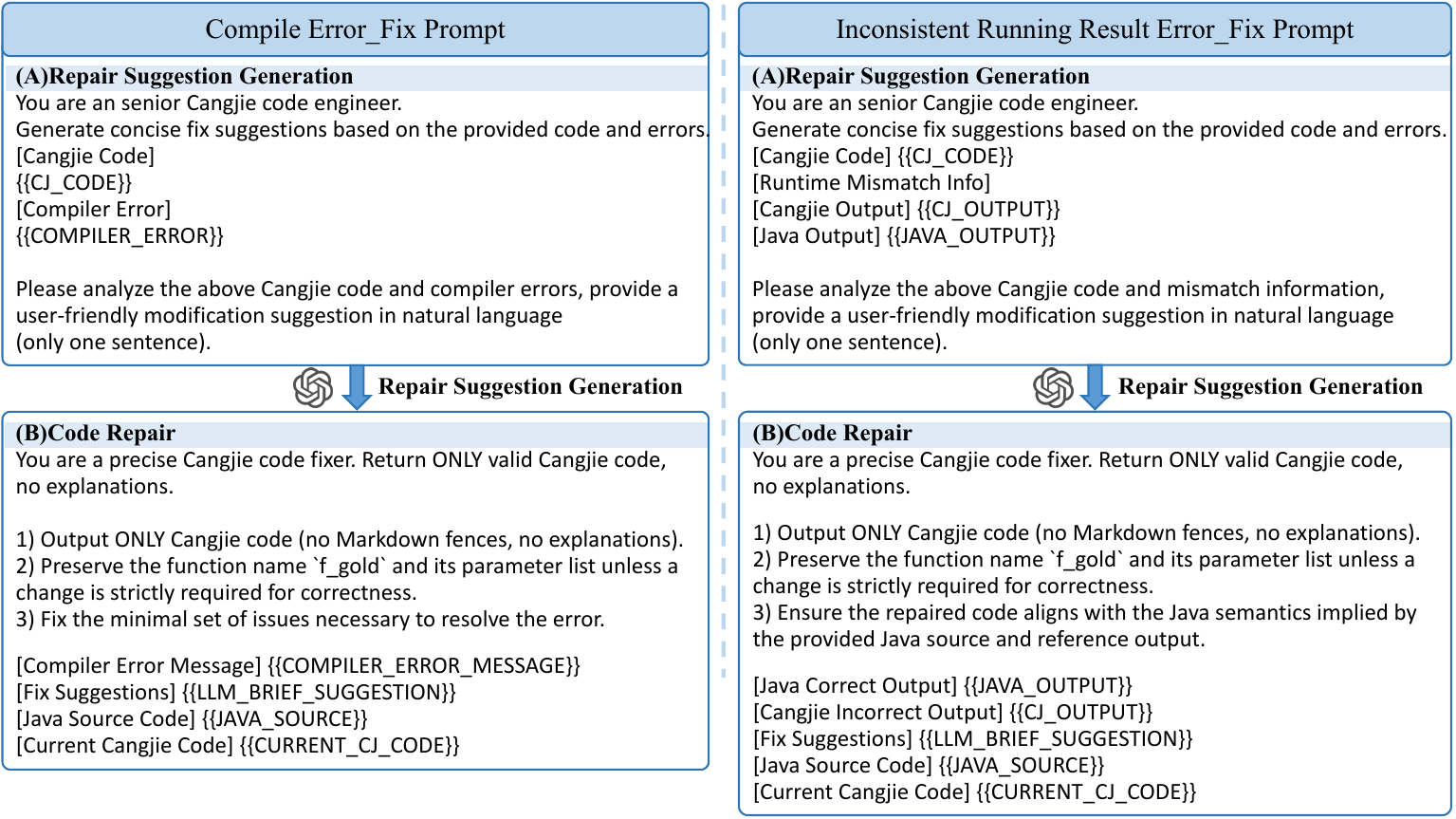}
    \caption{Prompt Templates for LLM-based Self-Analytic Repair with Compiler Feedback.}
    \label{fig:prompt-template}
\end{figure}

\textbf{LLM-based Self-Analysis Repair:} This module establishes an automated repair pipeline that sequentially executes code translation, compilation, and verification. Within this pipeline, we adopt an analyze-then-repair strategy to identify errors and apply targeted repairs. It first extracts the error messages from the compiler and testing results. Then, it designs a prompt template to combine the source Java code, the generated Cangjie code, and the error messages. Subsequently, the prompt is provided to the LLM to generate a repaired version of the Cangjie code. For the Cangjie code that is compiled successfully, we perform functional tests using a suite of 10 test cases \cite{cjTrans_2025}. Each source Java code in the test benchmark has a set of test cases, which are executed on the translated Cangjie code. By comparing the outputs, we identify failed samples with functional inconsistencies. Our approach then constructs a repair prompt using failed test input, Java-Cangjie output discrepancies, and related code snippets. 
The LLM is guided to explain the root cause and propose code changes before generating repaired code. The repair template is shown in Figure \ref{fig:prompt-template}. To avoid ineffective loops, our repair procedure tracks error types and key changes at each round. LLM-based self-analysis repair terminates when the error pattern stabilizes or the iteration limit is reached.

\begin{figure}
    \centering
    \includegraphics[width=1\linewidth]{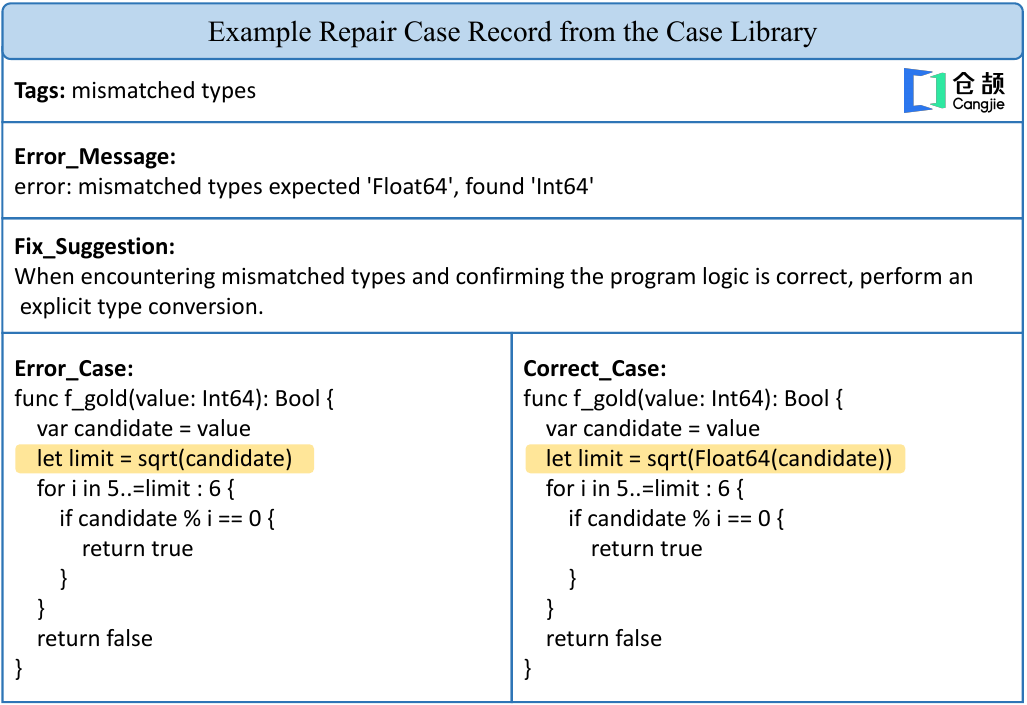}
    \caption{An Example of the error repair case in the error repair repository.}
    \label{fig:repair-case}
\end{figure}

\textbf{RAG-enhanced Repair:} We introduce RAG-enhanced repair with an error repair repository. Figure \ref{fig:repair-case} presents a specific example from the error repair repository. The repository stores a series of error repair cases with error tags, the error information, repair suggestions, code fragments triggering errors, and corrected code in a structured form. We execute large-scale Java-to-Cangjie translation on the training set and aggregate those cases successfully repaired by the LLM-based self-analysis to form the error repair repository, which contains 217 error repair cases.
These cases cover frequent and representative error patterns, and the repository can be continuously extended with new error-fix pairs to support evolving language features without retraining the model.


When compilation errors occur, our approach first retrieves similar error repair cases from the repository. Since compiler feedback typically contains explicit error types and keywords, we use a weighted, multi-dimensional similarity metric. To retrieve similar repair cases, our approach calculates a weighted similarity score based on six dimensions: error type features, keyword overlap, semantic similarity, code structure, character sequence, and edit distance. Specifically, the similarity calculation formula is defined as: 

\begin{equation}
Sim(Q, C_i) = \sum_{j=1}^{6} w_j \cdot S_j(Q, Ci)
\end{equation}
where $Q$ represents the query error, $C_i$ denotes the $i$-th candidate case, $S_j$ denotes the similarity function for the $j$-th dimension, and $w_j$ denotes the corresponding weight.

After we obtain similar error repair cases, our approach designs a prompt using the current compiler feedback, the retrieved similar error repair cases, and the code snippet with errors. The LLM can generate a repaired code snippet guided by the prompt.

\section{Experimental Setup}

We evaluate our approach by answering the following Research Questions (RQs).

\textbf{RQ1: What is the overall performance of our proposed approach?}

This RQ evaluates the comprehensive performance of our approach in terms of different evaluation metrics. Using the same test benchmark dataset, we compare our approach with two representative LLM-based baselines in the low-resource scenario.

\textbf{OptCodeTrans} \cite{OptTrans_2025} adopts a two-stage post-training strategy combining pre-training and instruction fine-tuning to improve translation performance across multiple programming paradigms.

\textbf{Compiler Feedback Translation (CFTTrans)} \cite{cjTrans_2025} first constructs initial parallel data using monolingual seed corpora and back-translation. It then applies compiler feedback to repair translated code, combined with KTO preference optimization for alignment training to improve translation performance.

\textbf{RQ2: What are the contributions of each stage in the training framework?}

We analyze the effectiveness of each training stage in this RQ, including syntactic knowledge pre-training, semantic monolingual instruction fine-tuning, and AST-aware parallel instruction fine-tuning. We conduct ablation studies by removing or simplifying each stage under the same experimental settings as the full approach. In these experiments, we disable the iterative error repair stage during inference to isolate the effect of training stages.

\textbf{RQ3: What are the effects of different repair strategies in the iterative error repair stage?}

This RQ evaluates the design of the repair stage, examining the differences and complementarity between LLM-based self-analysis repair and RAG-enhanced repair in improving the performance of code translation. Using the same benchmark dataset, we evaluate (i) only LLM-based self-analysis repair, (ii) only RAG-enhanced repair, and (iii) the full setting that combines both strategies.

\textbf{RQ4: How does our approach perform across different LLMs?}

We examine whether the same training pipeline leads to consistent performance for different LLMs in this RQ. We select several representative LLMs, executing the same training pipeline using the same training data and hyperparameters to compare their performance in code translation.

\subsection{Datasets}

\textbf{Syntactic Knowledge Dataset} is used in the CPT stage. We collect over 140 syntax chapters from the Cangjie development guides and standard API documentation. Then, we reformat them into structured syntactic knowledge to match the input format of CPT. The final constructed dataset contains 6779 samples and covers over 3400 core grammar concepts in Cangjie.

\textbf{Semantic Monolingual Dataset} is used in the monolingual instruction fine-tuning stage. We collect 11 repositories from the Cangjie community on GitCode, with a total of 1216 code snippets. After cleaning and filtering, we retain 1091 code snippets to construct the monolingual dataset. Statistical results show that the code snippets contain approximately 16 lines in length on average, with no code snippet containing fewer than 5 lines. To expand the scale of the monolingual dataset, we further introduce Cangjie code snippets from the parallel datasets and generate their corresponding semantic descriptions. The final dataset contains 3241 code snippets with semantic descriptions for training.

\textbf{AST-aware Parallel Dataset} is used in the parallel instruction fine-tuning stage. We use the publicly available OptCodeTrans parallel dataset \cite{OptTrans_2025} and obtain 2140 code snippets after cleaning and filtering. For each code snippet, we generate AST structure tokens from the Java code. We then assemble the code snippets and structure tokens using a predefined template to form the AST-aware parallel dataset.

\subsection{Selected Large Language Models}

We select representative LLMs as backbone models, including general-purpose LLMs and LLMs for code. To ensure experimental fairness, all the LLMs selected in this paper are released before the official release of Cangjie. Their base pre-training data do not include Cangjie-related syntax or API knowledge. Therefore, observed gains are attributed to the proposed training pipeline.
We prioritize mid-sized open-source models because the framework requires tokenizer extension and multi-stage fine-tuning, which are restricted in closed-source models.

\textbf{Llama-3.1-8B-Instruct} is an instruction-tuned model in the Llama 3.1 series. It is based on the Transformer architecture and is fine-tuned for instruction-conditioned generation. We use it to evaluate the effectiveness of our pipeline on a general-purpose backbone.

\textbf{Qwen2-7B-Instruct} \cite{qwen2} is an instruction-tuned model from the Qwen2 series. It performs well on language understanding, text generation, and code generation. We include it as a mid-sized model to test robustness across different general-purpose model families.

\textbf{StarCoder2-3B-Instruct} \cite{starcoder} is an open-source code LLM from the BigCode project. It is trained on large-scale code corpora from the Stack v2 across 17 programming languages. We use it to evaluate performance on code-specialized backbones and compare it with general instruction models in low-resource code translation.

\subsection{Evaluation Metrics}

To evaluate the performance of code translation, we employ four evaluation metrics as follows.

First, we use BLEU \cite{bleu} to measure the lexical overlap between the generated Cangjie code and the reference Cangjie code. Next, for functional evaluation, we also use Functional Equivalence (FE) \cite{FE_2024}. FE executes the source program and the generated program under the same inputs and checks whether the outputs are the same. If any test case fails to execute or the output are not exactly the same, the generated code is considered functionally inequivalent.

Then, we also calculate Compilation Success Rate (CSR) \cite{CSR_2025}, which measures the proportion of translated programs that are compiled successfully. Let $N_{compiled}$ denote the number of samples that successfully pass compilation, and let $N_{total}$ denote the total number of samples in the test set. CSR is defined as:
\begin{equation}
CSR = \frac{N_{compiled}}{N_{total}}
\end{equation}

Finally, to separate non-compileable errors from compilable-but-incorrect outputs, we introduce compilable Conditional Functional Equivalence (CFE). Let $N_{cf}$ denote the number of compilable samples that pass all tests. CFE is defined as:
\begin{equation}
CFE = \frac{N_{cf}}{N_{compiled}}
\end{equation}

CFE calculates the functional correctness proportion only on the set of samples that pass compilation, measuring the performance to maintain semantic consistency after satisfying compilation constraints. 

\subsection{Implementation and Default Parameters}

All experiments are completed under the same software and hardware environment to ensure reproducibility and fairness of experimental results. The experimental platform is equipped with two NVIDIA RTX 4090 GPUs to support the training and inference processes of LLMs. The operating system is Ubuntu 22.04, the experimental environment is built on Python 3.12, PyTorch 2.5.1, and CUDA 12.4. We use Llama-3.1-8B as the default LLM.

During inference, we use consistent decoding settings across all experiments. We adopt stochastic decoding with top\_p = 1.0 and temperature = 0. This setting can keep the generation stable while maintaining diversity, following common practice in code-related tasks \cite{codePlan_2024}. In the RAG-enhanced repair stage, we set the similarity threshold to 0.5.


\section{Experimental Results and Analysis}

\subsection{RQ1: What is the overall performance of our proposed approach?}

Table \ref{tb:rq1} reports the results of our approach and the baselines. Overall, our approach achieves the highest FE. This result suggests that, in the low-resource scenario, the proposed multi-stage training and iterative error repair pipeline for LLM improves the functional correctness. Specifically, our approach achieves FE of 63.64\%, which is 20.61\% higher than OptCodeTrans and 6.06\% higher than CFTTrans. In terms of text-level similarity, our approach also achieves the highest BLEU score. This indicates that the generated Cangjie code is similar to the reference Cangjie code.

In terms of CFE, our approach achieves 88.98\%, which is higher than the two baselines. This result indicates that our approach is more stable in terms of semantic preservation and structural alignment, which is consistent with our goal of improving translation robustness and functional correctness. Additionally, the CSR achieved by our approach is not the best, mainly due to the limited training data scale. In contrast, CFTTrans uses tens of thousands of Cangjie code snippets as training corpora, which improves the CSR. However, our training set scale used in this paper contains only approximately 3241 samples, with relatively limited coverage. Nevertheless, a higher CSR does not guarantee higher functional correctness. CFTTrans reports a much lower CFE than our approach, indicating that semantic deviations still exist even when the generated code can be compiled. In contrast, our approach achieves higher CFE, which suggests better functional consistency.

\begin{table}[t]
\caption{Overall performance comparison with baselines.}
\label{tb:rq1}
\begin{tabular}{ccccc}
\toprule
Approach                        & FE(\%)               & CSR(\%)              & CFE(\%)              & BLEU             \\ \hline
OptCodeTrans                  & 43.03          & 52.73          & 81.61          & 69.83          \\
\makecell{CFTTrans} & 57.58          & \textbf{79.39} & 72.52          & 73.64          \\
Our approach                    & \textbf{63.64} & 71.52          & \textbf{88.98} & \textbf{73.79} \\ \bottomrule
\end{tabular}
\end{table}

\begin{mybox}{Conclusion of RQ1}
The proposed approach achieves better results than the state-of-the-art approaches in code translation, which improves functional correctness in the low-resource scenario.
\end{mybox}

\subsection{RQ2: What are the contributions of each stage in the training framework?}

To evaluate the contribution of each training stage, we conduct an ablation study. Table \ref{tb:rq2} reports the results under different ablation settings. Specifically, \textit{only Parallel SFT} represents using only parallel dataset for instruction fine-tuning. This setting can reflect the most common paradigm in existing code translation. \textit{w/o CPT} removes syntactic knowledge continued pre-training. \textit{w/o SE-Mono SFT} retains the monolingual instruction fine-tuning but removes semantic enhancement signals. \textit{w/o AST-Parallel SFT} removes all AST-related content during parallel SFT and keeps only the instruction and source code as input.

The overall performance is notably reduced when using only Parallel SFT (FE = 36.97\%, CSR = 45.45\%). This indicates that limited parallel dataset is insufficient for learning both reliable Cangjie code generation and robust cross-language semantic alignment. Notably, this model still achieves a BLEU score of 68.37, while the FE and CSR scores are low. This inconsistency suggests that BLEU primarily reflects text-level similarity and cannot reliably measure functional correctness.

The CPT stage makes a clear contribution. Compared with the full approach, \textit{w/o CPT} leads to a continuous decline in performance: FE drops from 50.30\% to 43.03\%, and CSR drops from 55.76\% to 50.30\%. This indicates that CPT provides our appraoch with essential Cangjie knowledge and generation priors.

Semantic signals and AST structural information provide additional benefits beyond CPT. Compared to the full approach, removing semantic signals reduces FE by 4.85\% and CSR by 3.03\%, while decreasing CFE from 90.22\% to 86.21\%. These results suggest that semantic signals help our approach generate the target program whose behavior closely matches the source program. The decrease in CFE further indicates that instruction tuning with semantic signals improves the functional correctness of code translation. Removing the AST information also reduces performance. This suggests that AST constraints can help maintain structural alignment and reduce invalid structural outputs during parallel learning.

Overall, each training stage contributes to the final performance. The full achieves an FE of 50.30\% and a CSR of 55.76\%, representing improvements of 13.33\% and 10.31\% respectively compared to only Parallel SFT. These results indicate that multi-stage training improves both the ability to generate Cangjie code while preserving their functionalities.

\begin{table}[t]
\caption{Ablation study of the multi-stage training pipeline without inference-time repair.}
\label{tb:rq2}
\begin{tabular}{ccccc}
\toprule
Ablation Setting          & FE(\%)         & CSR(\%)        & CFE(\%)        & BLEU           \\ \hline
only Parallel SFT         & 36.97          & 45.45          & 81.33          & 68.37          \\
w/o CPT                   & 43.03          & 50.30          & 85.54          & 67.05          \\
w/o SE-Mono SFT           & 45.45          & 52.73          & 86.21          & 69.19          \\
w/o AST-Parallel SFT    & 46.67          & 53.94          & 86.52          & 70.13          \\
Our approach - w/o repair & \textbf{50.30} & \textbf{55.76} & \textbf{90.22} & \textbf{71.21} \\ \bottomrule
\end{tabular}
\end{table}

\begin{mybox}{Conclusion of RQ2}
Each stage positively contributes to the overall performance of our approach. These stages collectively enhance the compilation and functional correctness of low-resource code translation.
\end{mybox}

\subsection{RQ3: What are the effects of different repair strategies in the iterative error repair stage?}

To assess how different error repair strategies affect the performance of code translation, we conducted an ablation study using the same trained LLM and vary only the repair setting in inference. The overall results are shown in Table \ref{tb:rq3}. \textit{Only train} uses no repair strategy and serves as the baseline. \textit{Only train + LLM repair} uses only LLM-based self-analysis repair. \textit{Only train + RAG repair} applies only RAG-enhanced repair. In contrast, our full approach combines LLM-based self-analysis repair with RAG.

From Table \ref{tb:rq3}, after introducing repair strategies based on training, all metrics obtain stable gains. This suggests that iterative error repair can effectively improve the performance of code translation. Compared with \textit{only train}, \textit{only train + LLM} repair increases FE by 6.67\% and CSR by 7.88\%. Under the setting of using only RAG-enhanced repair, the improvement is larger: FE reaches 61.21\%, with an increase of 10.91\% compared to \textit{only train}, and CSR improves by 13.33\%. Compared with LLM-based self-analysis repair, RAG retrieves similar error repair cases and provides more specific suggestions. As a result, the repaired code is closer to the correct code. Furthermore, the complete framework achieves an FE of 63.64\%, a CSR of 71.52\%, and a BLEU score of 73.79, corresponding to improvements of 13.34\% and 15.76\% in FE and CSR, and a 2.58 increase in BLEU. Our approach also outperforms \textit{only train + RAG repair}, which suggests that the two strategies are complementary.

It should be noted that CFE slightly decreases after enabling error repair. In practice, the error repair step almost never breaks code that already compiles. Instead, it mainly converts previously non-compilable outputs into compilable ones without fixing the underlying semantic issues, which reduces the proportion of functionally correct code within the compilable set.


\begin{table}[t]
\caption{Effects of different repair strategies on code translation.}
\label{tb:rq3}
\begin{tabular}{ccccc}
\toprule
Method                & FE(\%)         & CSR(\%)        & CFE(\%)        & BLEU           \\ \hline
Only train            & 50.30          & 55.76          & \textbf{90.22} & 71.21          \\
Only train+LLM repair & 56.97          & 63.64          & 89.52          & 72.46          \\
Only train+RAG repair & 61.21          & 69.09          & 88.60          & 73.17          \\
Our approach          & \textbf{63.64} & \textbf{71.52} & 88.98          & \textbf{73.79} \\
\bottomrule
\end{tabular}
\end{table}

\begin{mybox}{Conclusion of RQ3}
Combining both repair strategies improves the correctness of translated code. RAG-enhanced repair shows a greater improvement, indicating that error repair examples are important in code translation.
\end{mybox}

\subsection{RQ4: How does our approach perform across different LLMs?}

\begin{figure}
    \centering
    \includegraphics[width=1\linewidth]{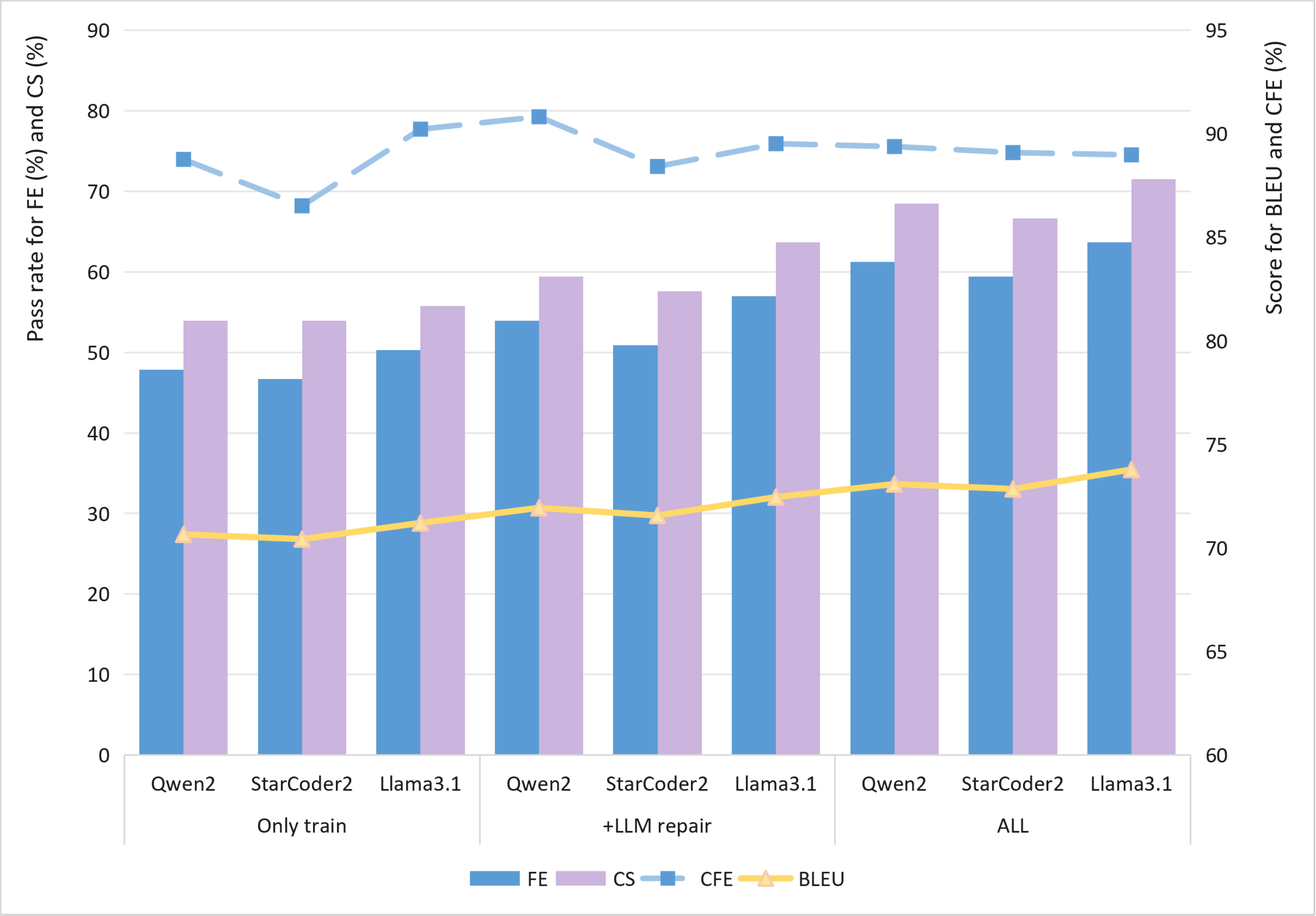}
    \caption{Code Translation Performance across Backbone LLMs under Different Settings.}
    \label{fig:pic-diff}
\end{figure}

To evaluate whether our approach is sensitive to the choice of LLMs, we conducted a comparative experiment with three representative LLMs, including two general LLMs and one LLM for code. Figure \ref{fig:pic-diff} presents a comprehensive comparison of code translation performance across different LLMs under various settings. Bar charts correspond to FE and CSR, uniformly mapped to the left axis. Lines correspond to CFE and BLEU, using the right axis. \textit{Only train} represents using only the training pipeline without the error repair stage, \textit{+LLM repair} adds LLM-based self-analysis repair, and ALL means the full approach.

The results show that our approach performs well and is consistent with different LLMs. Without any repair, our approach achieves different performances among LLMs, with FE ranging from 46.67\% to 50.30\%. After applying the full approach, each LLM gains a comparable improvement. For instance, qwen2-7b improves FE from 47.88\% to 61.21\%, StarCoder2-3B from 46.67\% to 59.39\%, and Llama3.1-8B from 50.30\% to 63.64\%. CSR respectively increases by 14.54\%, 12.73\%, and 15.76\% for the three LLMs. These trends suggest that the training and repair components generalize well across different LLMs. After enabling LLM-based self-analysis repair, both FE and CSR increase for every LLM. This indicates that compiler feedback provides a useful external signal that helps correct syntax and type errors, which are common when the LLM has limited knowledge of the target language.

Across the full setting, each LLM reaches its best overall performance. Meanwhile, CFE remains consistently high with only small variation across different LLMs, suggesting that higher compilation success does not come at the cost of semantic reliability.

\begin{mybox}{Conclusion of RQ4}
Our approach achieves comparable results across different LLMs, meaning that our approach is insensitive to the choice of LLMs.
\end{mybox}

\section{Threats to Validity}

\textbf{Internal Threats:} The LLM-generated syntactic knowledge dataset may contain minor errors. We manually verify and revise them before the pre-training of the LLM to improve its correctness. In addition, the testing results of the generated Cangjie code depend on the test coverage. We use established public benchmarks that provide ten test cases per task to reduce this bias.

\textbf{External Threats:} Our evaluation focuses on fragment-level translation rather than project-level migration, as our approach relies on general structure-aware cues rather than project-specific heuristics. Furthermore, the evaluation is limited to the Java-to-Cangjie language pair. However, our training components employ language-agnostic rules, and the consistent improvements observed across multiple LLM backbones provide preliminary evidence of generalizability.

\section{Related Work}

\subsection{Code Translation}

Early code translation approaches relied on hand-crafted heuristic rules \cite{ruleBad2_2014} or Statistical Machine Translation (SMT) \cite{SMTcode_2014,SMTbug_2016}, which struggled with cross-ecosystem structural and semantic consistency. Subsequently, deep learning introduced Neural Machine Translation (NMT) frameworks \cite{NMT_2014}, such as Tree-to-Tree models \cite{neural_Tree_2018} and unsupervised approaches like TransCoder \cite{Roziere_2020}, which used back-translation and monolingual data. Later work introduced intermediate representations to improve structural alignment \cite{roziere_2022}. However, these approaches still provide limited coverage for low-resource languages \cite{lowresICPC_2025}.


Recent work applies LLMs through pre-training and prompt designing. Yet zero-shot generation often contains semantic errors and fails to compile or run \cite{lost_2024}. As a result, research has moved to iterative methods with execution feedback \cite{unitrans_2024}, such as multi-stage post-training \cite{OptTrans_2025} and compiler-guided alignment \cite{cjTrans_2025}. Based on these work, we propose a multi-stage training and iterative error repair framework for the low-resource scenario in this study. 

\subsection{LLM-Based Error Repair}

With the rise of LLMs, Automated Program Repair (APR) initially focused on constrained generation for syntactically correct patches \cite{AlphaRepair_2022}. Recently, the focus has shifted toward feedback-driven repair.  
Frameworks such as ChatRepair demonstrate the efficacy of incorporating test failure information into the iterative generation to reduce ineffective attempts. In code translation, recent methodologies \cite{transRepair_2025,c2rust_2025} emphasize coupling LLM generation with compiler diagnostics and execution signals. These work collectively establish that one-shot generation is insufficient for complex code translation tasks, validating the necessity of iterative refinement mechanisms.

\section{Conclusion}


This paper proposes a multi-stage training and iterative error repair framework for code translation in low-resource ecosystems, combining a three-stage training pipeline with iterative error repair at inference time. 
Results show that this approach can significantly improve code translation capabilities on low-resource programming languages. Ablation studies also verify the contribution of each stage. Meanwhile, since Cangjie is an evolving language, the framework remains modular and can be easily adapted by updating the dataset and repository, ensuring stability across language updates.

\begin{acks}
This work was supported by the National Natural Science Foundation of China under Grant No. 62272225 and the Open Fund of the Shanghai Key Laboratory of Space-based Heterogeneous Network Collaborative Computing.
\end{acks}

\bibliographystyle{ACM-Reference-Format}
\bibliography{reference}


\begin{thebibliography}{48}


\ifx \showCODEN    \undefined \def \showCODEN     #1{\unskip}     \fi
\ifx \showISBNx    \undefined \def \showISBNx     #1{\unskip}     \fi
\ifx \showISBNxiii \undefined \def \showISBNxiii  #1{\unskip}     \fi
\ifx \showISSN     \undefined \def \showISSN      #1{\unskip}     \fi
\ifx \showLCCN     \undefined \def \showLCCN      #1{\unskip}     \fi
\ifx \shownote     \undefined \def \shownote      #1{#1}          \fi
\ifx \showarticletitle \undefined \def \showarticletitle #1{#1}   \fi
\ifx \showURL      \undefined \def \showURL       {\relax}        \fi
\providecommand\bibfield[2]{#2}
\providecommand\bibinfo[2]{#2}
\providecommand\natexlab[1]{#1}
\providecommand\showeprint[2][]{arXiv:#2}

\bibitem[Haugeland et~al\mbox{.}(2021)]%
        {migrate00_2021}
\bibfield{author}{\bibinfo{person}{Sindre~Grønstøl Haugeland}, \bibinfo{person}{Phu~H. Nguyen}, \bibinfo{person}{Hui Song}, {and} \bibinfo{person}{Franck Chauvel}.} \bibinfo{year}{2021}\natexlab{}.
\newblock \showarticletitle{Migrating Monoliths to Microservices-based Customizable Multi-tenant Cloud-native Apps}. In \bibinfo{booktitle}{\emph{2021 47th Euromicro Conference on Software Engineering and Advanced Applications (SEAA)}}. \bibinfo{pages}{170--177}.
\newblock
\href{https://doi.org/10.1109/SEAA53835.2021.00030}{doi:\nolinkurl{10.1109/SEAA53835.2021.00030}}


\bibitem[Krishna et~al\mbox{.}(2022)]%
        {migrate01_2022}
\bibfield{author}{\bibinfo{person}{Rahul Krishna}, \bibinfo{person}{Anup Kalia}, \bibinfo{person}{Saurabh Sinha}, \bibinfo{person}{Rachel Tzoref-Brill}, \bibinfo{person}{John Rofrano}, {and} \bibinfo{person}{Jin Xiao}.} \bibinfo{year}{2022}\natexlab{}.
\newblock \showarticletitle{Transforming monolithic applications to microservices with Mono2Micro}. In \bibinfo{booktitle}{\emph{Proceedings of the 36th IEEE/ACM International Conference on Automated Software Engineering}} (Melbourne, Australia) \emph{(\bibinfo{series}{ASE '21})}. \bibinfo{publisher}{IEEE Press}, \bibinfo{pages}{3}.
\newblock
\showISBNx{9781665403375}
\href{https://doi.org/10.1109/ASE51524.2021.9678851}{doi:\nolinkurl{10.1109/ASE51524.2021.9678851}}


\bibitem[Romani et~al\mbox{.}(2022)]%
        {migrate02_2022}
\bibfield{author}{\bibinfo{person}{Yamina Romani}, \bibinfo{person}{Okba Tibermacine}, {and} \bibinfo{person}{Chouki Tibermacine}.} \bibinfo{year}{2022}\natexlab{}.
\newblock \showarticletitle{Towards Migrating Legacy Software Systems to Microservice-based Architectures: a Data-Centric Process for Microservice Identification}. In \bibinfo{booktitle}{\emph{2022 IEEE 19th International Conference on Software Architecture Companion (ICSA-C)}}. \bibinfo{pages}{15--19}.
\newblock
\href{https://doi.org/10.1109/ICSA-C54293.2022.00010}{doi:\nolinkurl{10.1109/ICSA-C54293.2022.00010}}


\bibitem[Chunchu(2025)]%
        {migrate03_2025}
\bibfield{author}{\bibinfo{person}{Abhinav Chunchu}.} \bibinfo{year}{2025}\natexlab{}.
\newblock \showarticletitle{Generative AI-Driven Legacy System Modernization: Transforming Enterprise Infrastructure Through Automated Code Translation and Refactoring}.
\newblock \bibinfo{journal}{\emph{Journal of Computer Science and Technology Studies}} \bibinfo{volume}{7}, \bibinfo{number}{6} (\bibinfo{year}{2025}), \bibinfo{pages}{407--414}.
\newblock


\bibitem[Tao et~al\mbox{.}(2024)]%
        {howfar_2024}
\bibfield{author}{\bibinfo{person}{Qingxiao Tao}, \bibinfo{person}{Tingrui Yu}, \bibinfo{person}{Xiaodong Gu}, {and} \bibinfo{person}{Beijun Shen}.} \bibinfo{year}{2024}\natexlab{}.
\newblock \showarticletitle{Unraveling the Potential of Large Language Models in Code Translation: How Far are We?}. In \bibinfo{booktitle}{\emph{2024 31st Asia-Pacific Software Engineering Conference (APSEC)}}. \bibinfo{pages}{353--362}.
\newblock
\href{https://doi.org/10.1109/APSEC65559.2024.00046}{doi:\nolinkurl{10.1109/APSEC65559.2024.00046}}


\bibitem[Khyber~Sen(2017)]%
        {c2rust}
\bibfield{author}{\bibinfo{person}{Andrei~Homescu Khyber~Sen}.} \bibinfo{year}{2017}\natexlab{}.
\newblock \bibinfo{booktitle}{\emph{c2rust}}.
\newblock
\urldef\tempurl%
\url{https://github.com/immunant/c2rust}
\showURL{%
Retrieved January 15, 2026 from \tempurl}


\bibitem[Smirnov(2021)]%
        {cxgo}
\bibfield{author}{\bibinfo{person}{Denys Smirnov}.} \bibinfo{year}{2021}\natexlab{}.
\newblock \bibinfo{booktitle}{\emph{cxgo}}.
\newblock
\urldef\tempurl%
\url{https://github.com/gotranspile/cxgo}
\showURL{%
Retrieved January 15, 2026 from \tempurl}


\bibitem[Paul~Irwin(2016)]%
        {java2csharp}
\bibfield{author}{\bibinfo{person}{Maximilien~Noal Paul~Irwin, Vahid~Nasiri}.} \bibinfo{year}{2016}\natexlab{}.
\newblock \bibinfo{booktitle}{\emph{JavaToCSharp}}.
\newblock
\urldef\tempurl%
\url{https://github.com/paulirwin/JavaToCSharp}
\showURL{%
Retrieved January 15, 2026 from \tempurl}


\bibitem[Nguyen et~al\mbox{.}(2013)]%
        {ruleBad_2013}
\bibfield{author}{\bibinfo{person}{Anh~Tuan Nguyen}, \bibinfo{person}{Tung~Thanh Nguyen}, {and} \bibinfo{person}{Tien~N. Nguyen}.} \bibinfo{year}{2013}\natexlab{}.
\newblock \showarticletitle{Lexical statistical machine translation for language migration}. In \bibinfo{booktitle}{\emph{Proceedings of the 2013 9th Joint Meeting on Foundations of Software Engineering}} (Saint Petersburg, Russia) \emph{(\bibinfo{series}{ESEC/FSE 2013})}. \bibinfo{publisher}{Association for Computing Machinery}, \bibinfo{address}{New York, NY, USA}, \bibinfo{pages}{651–654}.
\newblock
\showISBNx{9781450322379}
\href{https://doi.org/10.1145/2491411.2494584}{doi:\nolinkurl{10.1145/2491411.2494584}}


\bibitem[Nguyen et~al\mbox{.}(2015)]%
        {statis_DaC_2015}
\bibfield{author}{\bibinfo{person}{Anh~Tuan Nguyen}, \bibinfo{person}{Tung~Thanh Nguyen}, {and} \bibinfo{person}{Tien~N. Nguyen}.} \bibinfo{year}{2015}\natexlab{}.
\newblock \showarticletitle{Divide-and-conquer approach for multi-phase statistical migration for source code}. In \bibinfo{booktitle}{\emph{Proceedings of the 30th IEEE/ACM International Conference on Automated Software Engineering}} (Lincoln, Nebraska) \emph{(\bibinfo{series}{ASE '15})}. \bibinfo{publisher}{IEEE Press}, \bibinfo{pages}{585–596}.
\newblock
\showISBNx{9781509000241}
\href{https://doi.org/10.1109/ASE.2015.74}{doi:\nolinkurl{10.1109/ASE.2015.74}}


\bibitem[Chen et~al\mbox{.}(2018)]%
        {neural_Tree_2018}
\bibfield{author}{\bibinfo{person}{Xinyun Chen}, \bibinfo{person}{Chang Liu}, {and} \bibinfo{person}{Dawn Song}.} \bibinfo{year}{2018}\natexlab{}.
\newblock \showarticletitle{Tree-to-tree neural networks for program translation}. In \bibinfo{booktitle}{\emph{Proceedings of the 32nd International Conference on Neural Information Processing Systems}} (Montr\'{e}al, Canada) \emph{(\bibinfo{series}{NIPS'18})}. \bibinfo{publisher}{Curran Associates Inc.}, \bibinfo{address}{Red Hook, NY, USA}, \bibinfo{pages}{2552–2562}.
\newblock


\bibitem[Szafraniec et~al\mbox{.}(2022)]%
        {roziere_2022}
\bibfield{author}{\bibinfo{person}{Marc Szafraniec}, \bibinfo{person}{Baptiste Roziere}, \bibinfo{person}{Hugh Leather}, \bibinfo{person}{Francois Charton}, \bibinfo{person}{Patrick Labatut}, {and} \bibinfo{person}{Gabriel Synnaeve}.} \bibinfo{year}{2022}\natexlab{}.
\newblock \showarticletitle{Code translation with compiler representations}.
\newblock \bibinfo{journal}{\emph{arXiv preprint arXiv:2207.03578}} (\bibinfo{year}{2022}).
\newblock


\bibitem[Yan et~al\mbox{.}(2023)]%
        {codetransocean_2023}
\bibfield{author}{\bibinfo{person}{Weixiang Yan}, \bibinfo{person}{Yuchen Tian}, \bibinfo{person}{Yunzhe Li}, \bibinfo{person}{Qian Chen}, {and} \bibinfo{person}{Wen Wang}.} \bibinfo{year}{2023}\natexlab{}.
\newblock \showarticletitle{{C}ode{T}rans{O}cean: A Comprehensive Multilingual Benchmark for Code Translation}. In \bibinfo{booktitle}{\emph{Findings of the Association for Computational Linguistics: EMNLP 2023}}, \bibfield{editor}{\bibinfo{person}{Houda Bouamor}, \bibinfo{person}{Juan Pino}, {and} \bibinfo{person}{Kalika Bali}} (Eds.). \bibinfo{publisher}{Association for Computational Linguistics}, \bibinfo{address}{Singapore}, \bibinfo{pages}{5067--5089}.
\newblock
\href{https://doi.org/10.18653/v1/2023.findings-emnlp.337}{doi:\nolinkurl{10.18653/v1/2023.findings-emnlp.337}}


\bibitem[Jana et~al\mbox{.}(2024)]%
        {cotran_2024}
\bibfield{author}{\bibinfo{person}{Prithwish Jana}, \bibinfo{person}{Piyush Jha}, \bibinfo{person}{Haoyang Ju}, \bibinfo{person}{Gautham Kishore}, \bibinfo{person}{Aryan Mahajan}, {and} \bibinfo{person}{Vijay Ganesh}.} \bibinfo{year}{2024}\natexlab{}.
\newblock \showarticletitle{{CoTran}: {An} {LLM}-{Based} {Code} {Translator} {Using} {Reinforcement} {Learning} with {Feedback} from {Compiler} and {Symbolic} {Execution}}.
\newblock In \bibinfo{booktitle}{\emph{Frontiers in {Artificial} {Intelligence} and {Applications}}}, \bibfield{editor}{\bibinfo{person}{Ulle Endriss}, \bibinfo{person}{Francisco~S. Melo}, \bibinfo{person}{Kerstin Bach}, \bibinfo{person}{Alberto Bugarín-Diz}, \bibinfo{person}{José~M. Alonso-Moral}, \bibinfo{person}{Senén Barro}, {and} \bibinfo{person}{Fredrik Heintz}} (Eds.). \bibinfo{publisher}{IOS Press}.
\newblock
\showISBNx{9781643685489}
\href{https://doi.org/10.3233/FAIA240968}{doi:\nolinkurl{10.3233/FAIA240968}}


\bibitem[Pan et~al\mbox{.}(2024)]%
        {lost_2024}
\bibfield{author}{\bibinfo{person}{Rangeet Pan}, \bibinfo{person}{Ali~Reza Ibrahimzada}, \bibinfo{person}{Rahul Krishna}, \bibinfo{person}{Divya Sankar}, \bibinfo{person}{Lambert~Pouguem Wassi}, \bibinfo{person}{Michele Merler}, \bibinfo{person}{Boris Sobolev}, \bibinfo{person}{Raju Pavuluri}, \bibinfo{person}{Saurabh Sinha}, {and} \bibinfo{person}{Reyhaneh Jabbarvand}.} \bibinfo{year}{2024}\natexlab{}.
\newblock \showarticletitle{Lost in Translation: A Study of Bugs Introduced by Large Language Models while Translating Code}. In \bibinfo{booktitle}{\emph{Proceedings of the IEEE/ACM 46th International Conference on Software Engineering}} (Lisbon, Portugal) \emph{(\bibinfo{series}{ICSE '24})}. \bibinfo{publisher}{Association for Computing Machinery}, \bibinfo{address}{New York, NY, USA}, Article \bibinfo{articleno}{82}, \bibinfo{numpages}{13}~pages.
\newblock
\showISBNx{9798400702174}
\href{https://doi.org/10.1145/3597503.3639226}{doi:\nolinkurl{10.1145/3597503.3639226}}


\bibitem[Yang et~al\mbox{.}(2024)]%
        {unitrans_2024}
\bibfield{author}{\bibinfo{person}{Zhen Yang}, \bibinfo{person}{Fang Liu}, \bibinfo{person}{Zhongxing Yu}, \bibinfo{person}{Jacky~Wai Keung}, \bibinfo{person}{Jia Li}, \bibinfo{person}{Shuo Liu}, \bibinfo{person}{Yifan Hong}, \bibinfo{person}{Xiaoxue Ma}, \bibinfo{person}{Zhi Jin}, {and} \bibinfo{person}{Ge Li}.} \bibinfo{year}{2024}\natexlab{}.
\newblock \showarticletitle{Exploring and Unleashing the Power of Large Language Models in Automated Code Translation}.
\newblock \bibinfo{journal}{\emph{Proc. ACM Softw. Eng.}} \bibinfo{volume}{1}, \bibinfo{number}{FSE}, Article \bibinfo{articleno}{71} (\bibinfo{date}{July} \bibinfo{year}{2024}), \bibinfo{numpages}{24}~pages.
\newblock
\href{https://doi.org/10.1145/3660778}{doi:\nolinkurl{10.1145/3660778}}


\bibitem[Yuan et~al\mbox{.}(2024)]%
        {transagent_2024}
\bibfield{author}{\bibinfo{person}{Zhiqiang Yuan}, \bibinfo{person}{Weitong Chen}, \bibinfo{person}{Hanlin Wang}, \bibinfo{person}{Kai Yu}, \bibinfo{person}{Xin Peng}, {and} \bibinfo{person}{Yiling Lou}.} \bibinfo{year}{2024}\natexlab{}.
\newblock \showarticletitle{Transagent: An llm-based multi-agent system for code translation}.
\newblock \bibinfo{journal}{\emph{arXiv preprint arXiv:2409.19894}} (\bibinfo{year}{2024}).
\newblock


\bibitem[Cassano et~al\mbox{.}(2024)]%
        {lowresLLM_2024}
\bibfield{author}{\bibinfo{person}{Federico Cassano}, \bibinfo{person}{John Gouwar}, \bibinfo{person}{Francesca Lucchetti}, \bibinfo{person}{Claire Schlesinger}, \bibinfo{person}{Anders Freeman}, \bibinfo{person}{Carolyn~Jane Anderson}, \bibinfo{person}{Molly~Q Feldman}, \bibinfo{person}{Michael Greenberg}, \bibinfo{person}{Abhinav Jangda}, {and} \bibinfo{person}{Arjun Guha}.} \bibinfo{year}{2024}\natexlab{}.
\newblock \showarticletitle{Knowledge Transfer from High-Resource to Low-Resource Programming Languages for Code LLMs}.
\newblock \bibinfo{journal}{\emph{Proc. ACM Program. Lang.}} \bibinfo{volume}{8}, \bibinfo{number}{OOPSLA2}, Article \bibinfo{articleno}{295} (\bibinfo{date}{Oct.} \bibinfo{year}{2024}), \bibinfo{numpages}{32}~pages.
\newblock
\href{https://doi.org/10.1145/3689735}{doi:\nolinkurl{10.1145/3689735}}


\bibitem[Roziere et~al\mbox{.}(2020)]%
        {Roziere_2020}
\bibfield{author}{\bibinfo{person}{Baptiste Roziere}, \bibinfo{person}{Marie-Anne Lachaux}, \bibinfo{person}{Lowik Chanussot}, {and} \bibinfo{person}{Guillaume Lample}.} \bibinfo{year}{2020}\natexlab{}.
\newblock \showarticletitle{Unsupervised translation of programming languages}. In \bibinfo{booktitle}{\emph{Proceedings of the 34th International Conference on Neural Information Processing Systems}} (Vancouver, BC, Canada) \emph{(\bibinfo{series}{NIPS '20})}. \bibinfo{publisher}{Curran Associates Inc.}, \bibinfo{address}{Red Hook, NY, USA}, Article \bibinfo{articleno}{1730}, \bibinfo{numpages}{11}~pages.
\newblock
\showISBNx{9781713829546}


\bibitem[Roziere et~al\mbox{.}(2021)]%
        {roziereST_2021}
\bibfield{author}{\bibinfo{person}{Baptiste Roziere}, \bibinfo{person}{Jie~M Zhang}, \bibinfo{person}{Francois Charton}, \bibinfo{person}{Mark Harman}, \bibinfo{person}{Gabriel Synnaeve}, {and} \bibinfo{person}{Guillaume Lample}.} \bibinfo{year}{2021}\natexlab{}.
\newblock \showarticletitle{Leveraging automated unit tests for unsupervised code translation}.
\newblock \bibinfo{journal}{\emph{arXiv preprint arXiv:2110.06773}} (\bibinfo{year}{2021}).
\newblock


\bibitem[Zhu et~al\mbox{.}(2024)]%
        {NoDataAug_2024}
\bibfield{author}{\bibinfo{person}{Ming Zhu}, \bibinfo{person}{Mohimenul Karim}, \bibinfo{person}{Ismini Lourentzou}, {and} \bibinfo{person}{Daphne Yao}.} \bibinfo{year}{2024}\natexlab{}.
\newblock \showarticletitle{Semi-Supervised Code Translation Overcoming the Scarcity of Parallel Code Data}. In \bibinfo{booktitle}{\emph{Proceedings of the 39th IEEE/ACM International Conference on Automated Software Engineering}} (Sacramento, CA, USA) \emph{(\bibinfo{series}{ASE '24})}. \bibinfo{publisher}{Association for Computing Machinery}, \bibinfo{address}{New York, NY, USA}, \bibinfo{pages}{1545–1556}.
\newblock
\showISBNx{9798400712487}
\href{https://doi.org/10.1145/3691620.3695524}{doi:\nolinkurl{10.1145/3691620.3695524}}


\bibitem[Wang et~al\mbox{.}(2025)]%
        {cjTrans_2025}
\bibfield{author}{\bibinfo{person}{Jun Wang}, \bibinfo{person}{Chenghao Su}, \bibinfo{person}{Yijie Ou}, \bibinfo{person}{Yanhui Li}, \bibinfo{person}{Jialiang Tan}, \bibinfo{person}{Lin Chen}, {and} \bibinfo{person}{Yuming Zhou}.} \bibinfo{year}{2025}\natexlab{}.
\newblock \showarticletitle{Translating to a Low-Resource Language with Compiler Feedback: A Case Study on Cangjie}.
\newblock \bibinfo{journal}{\emph{IEEE Transactions on Software Engineering}} \bibinfo{volume}{51}, \bibinfo{number}{9} (\bibinfo{year}{2025}), \bibinfo{pages}{2671--2692}.
\newblock
\href{https://doi.org/10.1109/TSE.2025.3594908}{doi:\nolinkurl{10.1109/TSE.2025.3594908}}


\bibitem[Liu et~al\mbox{.}(2023)]%
        {lowres_2023}
\bibfield{author}{\bibinfo{person}{Fang Liu}, \bibinfo{person}{Jia Li}, {and} \bibinfo{person}{Li Zhang}.} \bibinfo{year}{2023}\natexlab{}.
\newblock \showarticletitle{Syntax and Domain Aware Model for Unsupervised Program Translation}. In \bibinfo{booktitle}{\emph{Proceedings of the 45th International Conference on Software Engineering}} (Melbourne, Victoria, Australia) \emph{(\bibinfo{series}{ICSE '23})}. \bibinfo{publisher}{IEEE Press}, \bibinfo{pages}{755–767}.
\newblock
\showISBNx{9781665457019}
\href{https://doi.org/10.1109/ICSE48619.2023.00072}{doi:\nolinkurl{10.1109/ICSE48619.2023.00072}}


\bibitem[Shi et~al\mbox{.}(2025)]%
        {cptSurvey_2024}
\bibfield{author}{\bibinfo{person}{Haizhou Shi}, \bibinfo{person}{Zihao Xu}, \bibinfo{person}{Hengyi Wang}, \bibinfo{person}{Weiyi Qin}, \bibinfo{person}{Wenyuan Wang}, \bibinfo{person}{Yibin Wang}, \bibinfo{person}{Zifeng Wang}, \bibinfo{person}{Sayna Ebrahimi}, {and} \bibinfo{person}{Hao Wang}.} \bibinfo{year}{2025}\natexlab{}.
\newblock \showarticletitle{Continual Learning of Large Language Models: A Comprehensive Survey}.
\newblock \bibinfo{journal}{\emph{ACM Comput. Surv.}} \bibinfo{volume}{58}, \bibinfo{number}{5}, Article \bibinfo{articleno}{120} (\bibinfo{date}{Nov.} \bibinfo{year}{2025}), \bibinfo{numpages}{42}~pages.
\newblock
\showISSN{0360-0300}
\href{https://doi.org/10.1145/3735633}{doi:\nolinkurl{10.1145/3735633}}


\bibitem[Jana et~al\mbox{.}(2023)]%
        {cotran_2023}
\bibfield{author}{\bibinfo{person}{Prithwish Jana}, \bibinfo{person}{Piyush Jha}, \bibinfo{person}{Haoyang Ju}, \bibinfo{person}{Gautham Kishore}, \bibinfo{person}{Aryan Mahajan}, {and} \bibinfo{person}{Vijay Ganesh}.} \bibinfo{year}{2023}\natexlab{}.
\newblock \showarticletitle{Cotran: An llm-based code translator using reinforcement learning with feedback from compiler and symbolic execution}.
\newblock \bibinfo{journal}{\emph{Proceedings of the 27th European Conference on Artificial Intelligence}}  \bibinfo{volume}{392} (\bibinfo{year}{2023}), \bibinfo{pages}{4011--4018}.
\newblock
\href{https://doi.org/10.3233/FAIA240968}{doi:\nolinkurl{10.3233/FAIA240968}}


\bibitem[Xin(2017)]%
        {fixBug1_2017}
\bibfield{author}{\bibinfo{person}{Qi Xin}.} \bibinfo{year}{2017}\natexlab{}.
\newblock \showarticletitle{Towards addressing the patch overfitting problem}. In \bibinfo{booktitle}{\emph{Proceedings of the 39th International Conference on Software Engineering Companion}} (Buenos Aires, Argentina) \emph{(\bibinfo{series}{ICSE-C '17})}. \bibinfo{publisher}{IEEE Press}, \bibinfo{pages}{489–490}.
\newblock
\showISBNx{9781538615898}
\href{https://doi.org/10.1109/ICSE-C.2017.42}{doi:\nolinkurl{10.1109/ICSE-C.2017.42}}


\bibitem[Parasaram et~al\mbox{.}(2022)]%
        {fixBug2_2022}
\bibfield{author}{\bibinfo{person}{Nikhil Parasaram}, \bibinfo{person}{Earl~T. Barr}, {and} \bibinfo{person}{Sergey Mechtaev}.} \bibinfo{year}{2022}\natexlab{}.
\newblock \showarticletitle{Trident: Controlling Side Effects in Automated Program Repair}.
\newblock \bibinfo{journal}{\emph{IEEE Transactions on Software Engineering}} \bibinfo{volume}{48}, \bibinfo{number}{12} (\bibinfo{year}{2022}), \bibinfo{pages}{4717--4732}.
\newblock
\href{https://doi.org/10.1109/TSE.2021.3124323}{doi:\nolinkurl{10.1109/TSE.2021.3124323}}


\bibitem[Xie et~al\mbox{.}(2023)]%
        {dataAug_2023}
\bibfield{author}{\bibinfo{person}{Yiqing Xie}, \bibinfo{person}{Atharva Naik}, \bibinfo{person}{Daniel Fried}, {and} \bibinfo{person}{Carolyn Rose}.} \bibinfo{year}{2023}\natexlab{}.
\newblock \showarticletitle{Data Augmentation for Code Translation with Comparable Corpora and Multiple References}. In \bibinfo{booktitle}{\emph{The 2023 Conference on Empirical Methods in Natural Language Processing}}.
\newblock
\urldef\tempurl%
\url{https://openreview.net/forum?id=8NA76tz7Jj}
\showURL{%
\tempurl}


\bibitem[Wong et~al\mbox{.}(2025)]%
        {motivate_2025}
\bibfield{author}{\bibinfo{person}{Kyle Wong}, \bibinfo{person}{Alfonso Amayuelas}, \bibinfo{person}{Liangming Pan}, {and} \bibinfo{person}{William~Yang Wang}.} \bibinfo{year}{2025}\natexlab{}.
\newblock \showarticletitle{Investigating the transferability of code repair for low-resource programming languages}. In \bibinfo{booktitle}{\emph{Findings of the Association for Computational Linguistics: NAACL 2025}}. \bibinfo{pages}{3410--3432}.
\newblock


\bibitem[Hu et~al\mbox{.}(2022)]%
        {lora_2022}
\bibfield{author}{\bibinfo{person}{Edward~J Hu}, \bibinfo{person}{Yelong Shen}, \bibinfo{person}{Phillip Wallis}, \bibinfo{person}{Zeyuan Allen-Zhu}, \bibinfo{person}{Yuanzhi Li}, \bibinfo{person}{Shean Wang}, \bibinfo{person}{Lu Wang}, \bibinfo{person}{Weizhu Chen}, {et~al\mbox{.}}} \bibinfo{year}{2022}\natexlab{}.
\newblock \showarticletitle{Lora: Low-rank adaptation of large language models.}
\newblock \bibinfo{journal}{\emph{ICLR}} \bibinfo{volume}{1}, \bibinfo{number}{2} (\bibinfo{year}{2022}), \bibinfo{pages}{3}.
\newblock


\bibitem[Baltaji et~al\mbox{.}(2025)]%
        {LLMsemantic_2025}
\bibfield{author}{\bibinfo{person}{Razan Baltaji}, \bibinfo{person}{Saurabh Pujar}, \bibinfo{person}{Martin Hirzel}, \bibinfo{person}{Louis Mandel}, \bibinfo{person}{Luca Buratti}, {and} \bibinfo{person}{Lav~R. Varshney}.} \bibinfo{year}{2025}\natexlab{}.
\newblock \showarticletitle{Cross-lingual Transfer in Programming Languages: An Extensive Empirical Study}.
\newblock \bibinfo{journal}{\emph{Transactions on Machine Learning Research}} (\bibinfo{year}{2025}).
\newblock
\showISSN{2835-8856}
\urldef\tempurl%
\url{https://openreview.net/forum?id=1PRBHKgQVM}
\showURL{%
\tempurl}


\bibitem[Haldar and Hockenmaier(2024)]%
        {LLMsemantic2_2024}
\bibfield{author}{\bibinfo{person}{Rajarshi Haldar} {and} \bibinfo{person}{Julia Hockenmaier}.} \bibinfo{year}{2024}\natexlab{}.
\newblock \showarticletitle{Analyzing the performance of large language models on code summarization}.
\newblock \bibinfo{journal}{\emph{arXiv preprint arXiv:2404.08018}} (\bibinfo{year}{2024}).
\newblock


\bibitem[Max~Brunsfeld(2013)]%
        {treesitter}
\bibfield{author}{\bibinfo{person}{Andrew~Hlynskyi Max~Brunsfeld, Amaan~Qureshi}.} \bibinfo{year}{2013}\natexlab{}.
\newblock \bibinfo{booktitle}{\emph{tree-sitter}}.
\newblock
\urldef\tempurl%
\url{https://github.com/tree-sitter/tree-sitter}
\showURL{%
Retrieved January 15, 2026 from \tempurl}


\bibitem[Lin et~al\mbox{.}(2025)]%
        {OptTrans_2025}
\bibfield{author}{\bibinfo{person}{Jianbo Lin}, \bibinfo{person}{Yi Shen}, \bibinfo{person}{Chuanyi Li}, \bibinfo{person}{Changan Niu}, {and} \bibinfo{person}{Bin Luo}.} \bibinfo{year}{2025}\natexlab{}.
\newblock \showarticletitle{OptCodeTrans: Boost LLMs on Low-Resource Programming Language Translation}.
\newblock \bibinfo{journal}{\emph{2025 IEEE/ACM Second International Conference on AI Foundation Models and Software Engineering (Forge)}} (\bibinfo{year}{2025}), \bibinfo{pages}{67--72}.
\newblock
\urldef\tempurl%
\url{https://api.semanticscholar.org/CorpusID:279800367}
\showURL{%
\tempurl}


\bibitem[Team et~al\mbox{.}(2024)]%
        {qwen2}
\bibfield{author}{\bibinfo{person}{Qwen Team} {et~al\mbox{.}}} \bibinfo{year}{2024}\natexlab{}.
\newblock \showarticletitle{Qwen2 technical report}.
\newblock \bibinfo{journal}{\emph{arXiv preprint arXiv:2407.10671}} \bibinfo{volume}{2}, \bibinfo{number}{3} (\bibinfo{year}{2024}).
\newblock


\bibitem[Lozhkov et~al\mbox{.}(2024)]%
        {starcoder}
\bibfield{author}{\bibinfo{person}{Anton Lozhkov}, \bibinfo{person}{Raymond Li}, \bibinfo{person}{Loubna~Ben Allal}, \bibinfo{person}{Federico Cassano}, \bibinfo{person}{Joel Lamy-Poirier}, \bibinfo{person}{Nouamane Tazi}, \bibinfo{person}{Ao Tang}, \bibinfo{person}{Dmytro Pykhtar}, \bibinfo{person}{Jiawei Liu}, \bibinfo{person}{Yuxiang Wei}, {et~al\mbox{.}}} \bibinfo{year}{2024}\natexlab{}.
\newblock \showarticletitle{Starcoder 2 and the stack v2: The next generation}.
\newblock \bibinfo{journal}{\emph{arXiv preprint arXiv:2402.19173}} (\bibinfo{year}{2024}).
\newblock


\bibitem[Papineni et~al\mbox{.}(2002)]%
        {bleu}
\bibfield{author}{\bibinfo{person}{Kishore Papineni}, \bibinfo{person}{Salim Roukos}, \bibinfo{person}{Todd Ward}, {and} \bibinfo{person}{Wei-Jing Zhu}.} \bibinfo{year}{2002}\natexlab{}.
\newblock \showarticletitle{BLEU: a method for automatic evaluation of machine translation}. In \bibinfo{booktitle}{\emph{Proceedings of the 40th Annual Meeting on Association for Computational Linguistics}} (Philadelphia, Pennsylvania) \emph{(\bibinfo{series}{ACL '02})}. \bibinfo{publisher}{Association for Computational Linguistics}, \bibinfo{address}{USA}, \bibinfo{pages}{311–318}.
\newblock
\href{https://doi.org/10.3115/1073083.1073135}{doi:\nolinkurl{10.3115/1073083.1073135}}


\bibitem[Xue et~al\mbox{.}(2024)]%
        {FE_2024}
\bibfield{author}{\bibinfo{person}{Min Xue}, \bibinfo{person}{Artur Andrzejak}, {and} \bibinfo{person}{Marla Leuther}.} \bibinfo{year}{2024}\natexlab{}.
\newblock \showarticletitle{An interpretable error correction method for enhancing code-to-code translation}. In \bibinfo{booktitle}{\emph{The Twelfth International Conference on Learning Representations}}.
\newblock
\urldef\tempurl%
\url{https://openreview.net/forum?id=fVxIEHGnVT}
\showURL{%
\tempurl}


\bibitem[Xue et~al\mbox{.}(2025)]%
        {CSR_2025}
\bibfield{author}{\bibinfo{person}{Pengyu Xue}, \bibinfo{person}{Linhao Wu}, \bibinfo{person}{Zhen Yang}, \bibinfo{person}{Chengyi Wang}, \bibinfo{person}{Xiang Li}, \bibinfo{person}{Yuxiang Zhang}, \bibinfo{person}{Jia Li}, \bibinfo{person}{Ruikai Jin}, \bibinfo{person}{Yifei Pei}, \bibinfo{person}{Zhaoyan Shen}, {et~al\mbox{.}}} \bibinfo{year}{2025}\natexlab{}.
\newblock \showarticletitle{ClassEval-T: Evaluating Large Language Models in Class-Level Code Translation}.
\newblock \bibinfo{journal}{\emph{Proceedings of the ACM on Software Engineering}} \bibinfo{volume}{2}, \bibinfo{number}{ISSTA} (\bibinfo{year}{2025}), \bibinfo{pages}{1421--1444}.
\newblock


\bibitem[Bairi et~al\mbox{.}(2024)]%
        {codePlan_2024}
\bibfield{author}{\bibinfo{person}{Ramakrishna Bairi}, \bibinfo{person}{Atharv Sonwane}, \bibinfo{person}{Aditya Kanade}, \bibinfo{person}{Vageesh~D. C.}, \bibinfo{person}{Arun Iyer}, \bibinfo{person}{Suresh Parthasarathy}, \bibinfo{person}{Sriram Rajamani}, \bibinfo{person}{B. Ashok}, {and} \bibinfo{person}{Shashank Shet}.} \bibinfo{year}{2024}\natexlab{}.
\newblock \showarticletitle{CodePlan: Repository-Level Coding using LLMs and Planning}.
\newblock \bibinfo{journal}{\emph{Proc. ACM Softw. Eng.}} \bibinfo{volume}{1}, \bibinfo{number}{FSE}, Article \bibinfo{articleno}{31} (\bibinfo{date}{July} \bibinfo{year}{2024}), \bibinfo{numpages}{24}~pages.
\newblock
\href{https://doi.org/10.1145/3643757}{doi:\nolinkurl{10.1145/3643757}}


\bibitem[Karaivanov et~al\mbox{.}(2014)]%
        {ruleBad2_2014}
\bibfield{author}{\bibinfo{person}{Svetoslav Karaivanov}, \bibinfo{person}{Veselin Raychev}, {and} \bibinfo{person}{Martin Vechev}.} \bibinfo{year}{2014}\natexlab{}.
\newblock \showarticletitle{Phrase-Based Statistical Translation of Programming Languages}. In \bibinfo{booktitle}{\emph{Proceedings of the 2014 ACM International Symposium on New Ideas, New Paradigms, and Reflections on Programming \& Software}} (Portland, Oregon, USA) \emph{(\bibinfo{series}{Onward! 2014})}. \bibinfo{publisher}{Association for Computing Machinery}, \bibinfo{address}{New York, NY, USA}, \bibinfo{pages}{173–184}.
\newblock
\showISBNx{9781450332101}
\href{https://doi.org/10.1145/2661136.2661148}{doi:\nolinkurl{10.1145/2661136.2661148}}


\bibitem[Nguyen et~al\mbox{.}(2014)]%
        {SMTcode_2014}
\bibfield{author}{\bibinfo{person}{Anh~Tuan Nguyen}, \bibinfo{person}{Tung~Thanh Nguyen}, {and} \bibinfo{person}{Tien~N. Nguyen}.} \bibinfo{year}{2014}\natexlab{}.
\newblock \showarticletitle{Migrating code with statistical machine translation} \emph{(\bibinfo{series}{ICSE Companion 2014})}. \bibinfo{publisher}{Association for Computing Machinery}, \bibinfo{address}{New York, NY, USA}, \bibinfo{pages}{544–547}.
\newblock
\showISBNx{9781450327688}
\href{https://doi.org/10.1145/2591062.2591072}{doi:\nolinkurl{10.1145/2591062.2591072}}


\bibitem[Nguyen(2016)]%
        {SMTbug_2016}
\bibfield{author}{\bibinfo{person}{Tien~N. Nguyen}.} \bibinfo{year}{2016}\natexlab{}.
\newblock \showarticletitle{Code migration with statistical machine translation} \emph{(\bibinfo{series}{SoftwareMining 2016})}. \bibinfo{publisher}{Association for Computing Machinery}, \bibinfo{address}{New York, NY, USA}, \bibinfo{pages}{2}.
\newblock
\showISBNx{9781450345118}
\href{https://doi.org/10.1145/2975961.2990477}{doi:\nolinkurl{10.1145/2975961.2990477}}


\bibitem[Bahdanau et~al\mbox{.}(2014)]%
        {NMT_2014}
\bibfield{author}{\bibinfo{person}{Dzmitry Bahdanau}, \bibinfo{person}{Kyunghyun Cho}, {and} \bibinfo{person}{Yoshua Bengio}.} \bibinfo{year}{2014}\natexlab{}.
\newblock \showarticletitle{Neural machine translation by jointly learning to align and translate}.
\newblock \bibinfo{journal}{\emph{arXiv preprint arXiv:1409.0473}} (\bibinfo{year}{2014}).
\newblock


\bibitem[Giagnorio et~al\mbox{.}(2025)]%
        {lowresICPC_2025}
\bibfield{author}{\bibinfo{person}{Alessandro Giagnorio}, \bibinfo{person}{Alberto Martin-Lopez}, {and} \bibinfo{person}{Gabriele Bavota}.} \bibinfo{year}{2025}\natexlab{}.
\newblock \showarticletitle{{ Enhancing Code Generation for Low-Resource Languages: No Silver Bullet }}. In \bibinfo{booktitle}{\emph{2025 IEEE/ACM 33rd International Conference on Program Comprehension (ICPC)}}. \bibinfo{publisher}{IEEE Computer Society}, \bibinfo{address}{Los Alamitos, CA, USA}, \bibinfo{pages}{478--488}.
\newblock
\href{https://doi.org/10.1109/ICPC66645.2025.00058}{doi:\nolinkurl{10.1109/ICPC66645.2025.00058}}


\bibitem[Xia and Zhang(2022)]%
        {AlphaRepair_2022}
\bibfield{author}{\bibinfo{person}{Chunqiu~Steven Xia} {and} \bibinfo{person}{Lingming Zhang}.} \bibinfo{year}{2022}\natexlab{}.
\newblock \showarticletitle{Less training, more repairing please: revisiting automated program repair via zero-shot learning}. In \bibinfo{booktitle}{\emph{Proceedings of the 30th ACM Joint European Software Engineering Conference and Symposium on the Foundations of Software Engineering}} (Singapore, Singapore) \emph{(\bibinfo{series}{ESEC/FSE 2022})}. \bibinfo{publisher}{Association for Computing Machinery}, \bibinfo{address}{New York, NY, USA}, \bibinfo{pages}{959–971}.
\newblock
\showISBNx{9781450394130}
\href{https://doi.org/10.1145/3540250.3549101}{doi:\nolinkurl{10.1145/3540250.3549101}}


\bibitem[Zhou et~al\mbox{.}(2025)]%
        {transRepair_2025}
\bibfield{author}{\bibinfo{person}{Bo Zhou}, \bibinfo{person}{Jiaqi Shi}, \bibinfo{person}{Ying Wang}, \bibinfo{person}{Li Li}, \bibinfo{person}{Tsz~On Li}, \bibinfo{person}{Hai Yu}, {and} \bibinfo{person}{Zhiliang Zhu}.} \bibinfo{year}{2025}\natexlab{}.
\newblock \showarticletitle{Porting Software Libraries to OpenHarmony: Transitioning from TypeScript or JavaScript to ArkTS}.
\newblock \bibinfo{journal}{\emph{Proc. ACM Softw. Eng.}} \bibinfo{volume}{2}, \bibinfo{number}{ISSTA}, Article \bibinfo{articleno}{ISSTA064} (\bibinfo{date}{June} \bibinfo{year}{2025}), \bibinfo{numpages}{22}~pages.
\newblock
\href{https://doi.org/10.1145/3728941}{doi:\nolinkurl{10.1145/3728941}}


\bibitem[Hong and Ryu(2025)]%
        {c2rust_2025}
\bibfield{author}{\bibinfo{person}{Jaemin Hong} {and} \bibinfo{person}{Sukyoung Ryu}.} \bibinfo{year}{2025}\natexlab{}.
\newblock \showarticletitle{Type-migrating C-to-Rust translation using a large language model}.
\newblock \bibinfo{journal}{\emph{Empirical Software Engineering}} \bibinfo{volume}{30}, \bibinfo{number}{1} (\bibinfo{year}{2025}), \bibinfo{pages}{3}.
\newblock


\end{thebibliography}

\end{document}